\documentclass{emulateapj}

\makeatletter
\newcommand\tsb[1]{\@textsubscript{\selectfont#1}}
\def\@textsubscript#1{{\m@th\ensuremath{_{\mbox{\fontsize\sf@size\z@#1}}}}}
\newcommand\tsp[1]{\@textsuperscript{\selectfont#1}}
\def\@textsuperscript#1{{\m@th\ensuremath{^{\mbox{\fontsize\sf@size\z@#1}}}}}

\providecommand{\e}[1]{\ensuremath{\times 10^{#1}}}
\providecommand{\lsr}[1]{\textit{V}\tsb{LSR}}
\providecommand{\lsolar}[1]{#1 \textit{L}\tsb{$\odot$}}
\providecommand{\jfour}{4\tsb{13}-4\tsb{14}}
\providecommand{\jthree}{3\tsb{12}-3\tsb{13}}
\providecommand{\jone}{1\tsb{10}-1\tsb{11}}
\providecommand{\jtwo}{2\tsb{11}-2\tsb{12}}
\providecommand{\jfive}{5\tsb{14}-5\tsb{15}}
\providecommand{\kms}{km s\tsp{-1}}
\providecommand{\pms}[2]{$^{+#1}_{-#2}$}
\providecommand{\f}{H\tsb{2}CO}
\providecommand{\cm}{cm\tsp{-3}}

\begin{document}

\title{Formaldehyde Densitometry of Galactic Star-Forming Regions Using The \f{} \jthree{} and \jfour{} Transitions}
    
\author{Patrick I. McCauley, Jeffrey G. Mangum, and Alwyn Wootten}
\affil{National Radio Astronomy Observatory, 520 Edgemont Road,
  Charlottesville, VA 22903} 
	
%\shorttitle{\f{} \texit{J} = 3 and 4 \textit{K}-Doublet Densitometry}
%\shortauthors{McCauley, Mangum, and Wootten}
    
\begin{abstract}
	
We present Green Bank Telescope (GBT) observations of the \jthree{}
(29 GHz) and \jfour{} (48 GHz) transitions of the \f{} molecule toward 
a sample of 23 well-studied star-forming regions. Analysis of  
the relative intensities of these transitions can be used to reliably \textit{measure} the 
densities of molecular cores. Adopting kinetic temperatures from the 
literature, we have employed a Large Velocity Gradient (LVG) model to 
derive the average hydrogen number density [\textit{n}(H\tsb{2})] 
within a 16$\arcsec$ beam toward each source. Densities in the range 
of 10\tsp{5.5}--10\tsp{6.5} \cm{} and ortho-formaldehyde column 
densities per unit line width between 10\tsp{13.5} and 10\tsp{14.5} cm\tsp{-2} (\kms{})\tsp{-1} are found for most objects, 
in general agreement with existing measurements.
A detailed analysis of the advantages and limitations to this densitometry technique is also presented. 
We find that \f{} \jthree{}/\jfour{} densitometry proves to be best suited to objects 
with \textit{T}\tsb{K} $\gtrsim$ 100 K, above which the \f{} LVG models 
become relatively independent of kinetic temperature. This study represents 
the first detection of these \f{} \textit{K}-doublet transitions in all but one object in our sample. The ease with 
which these transitions were detected, coupled with their unique sensitivity 
to spatial density, make them excellent monitors of density in molecular 
clouds for future experiments. We also report the detection of the 
9\tsb{2}--8\tsb{1} A\tsp{-} (29 GHz) transition of CH\tsb{3}OH toward 6 sources. 

\end{abstract}


\section{Introduction}
\label{intro}

A thorough understanding of star formation is predicated on a 
knowledge of the physical conditions that surround the process 
at each of its stages. The determination of these properties has proven to be a challenging endeavour 
for the early phases of star formation because molecular hydrogen (H\tsb{2}), the 
primary constituent of developing stars, cannot be measured 
directly. Instead, the excitation properties of trace molecules must 
be used to infer the conditions within the greater cloud. Thus, all 
measurements of density in molecular clouds are subject to the biases 
of the chosen tracer. Even more problematic are virial density 
estimates determined via inferred mass and source size due to the 
uncertainties inherent in both 	quantities. Formaldehyde (\f{}) is uniquely 
sensitive to spatial density and an ideal probe of molecular cloud 
cores for a number of reasons that are discussed in \S\ref{h2coProbe}.

The \jone{} (5 GHz) and \jtwo{} (14 GHz) \textit{K}-doublet 
transitions of \f{} have previously been used to measure density 
in a number of studies of galactic 
(e.g. \citealt{Hen80,DG87,Tur89,Zyl92}) and extragalactic \citep{Man08} star formation regions.
Our work employs an essentially identical strategy using the \jthree{} 
(29 GHz) and \jfour{} (48 GHz) transitions, which offer a few advantages 
over their lower excitation counterparts. Principally, the higher 
frequencies translate to smaller single-dish beam sizes (GBT beam sizes of 26$\arcsec$ 
and 16$\arcsec$ for the $J=3$ and $J=4$ transitions, respectively, as 
opposed to 153$\arcsec$ and 51$\arcsec$ for $J=1$ and $J=2$ \textit{K}-doublets). 
The added spatial resolution is important when 
considering that stars form in spatially compact regions within molecular 
clouds and thus large beam sizes may serve to dilute the areas of interest. 
This point is compounded by the fact that the lower excitation transitions 
are by their nature often more sensitive to spatial densities and kinetic 
temperatures lower than those of interest in studies of star-forming molecular cores. 
The \jthree{} and \jfour{} transitions are therefore more efficient probes 
of spatial density in this context.

Few studies of the $J=3$ and 4 \textit{K}-doublet transitions have been made due 
partially to their relatively low intensities and high centimeter-wave frequencies. 
The $J=3$ transition was first detected by \citet{Wel70} in absorption toward the 
radio continuum source Sgr B2. Subsequent experiments were conducted 
toward the brightest source in our sample, Orion-KL \citep{Wil80,MB80,Bas85}, 
after which exploration of this transition seemingly ends. We could find no previous 
measurements of the \jfour{} transition, whose yet higher frequency and 
weaker intensity pose additional observational difficulties.

Since our ability to detect these transitions and the reliability of deriving density 
measurements from them was uncertain, a sample of very bright and well-studied 
objects were chosen (Table~\ref{tab:sources}). Strong detections of both transitions proved to 
be fairly easy to obtain, requiring an average of 17 min of integration time, and the 
resulting spatial density measurements (10\tsp{5.5}--10\tsp{6.5} \cm{}) 
are consistent with what is known for molecular cores. These results are 
encouraging for the prospect of future experiments.

\begin{deluxetable*}{l r r c r}
\tabletypesize{\scriptsize}
\tablecolumns{5} 
\tablewidth{0pc} 
\tablecaption{Source Positions} 
\tablehead{ 
\colhead{} & 
\colhead{} & 
\colhead{} & 
\colhead{Distance} & 
\colhead{\lsr{}} 
\\ 
\colhead{Source} & 
\colhead{$\alpha$(J2000.00)} & 
\colhead{$\delta$(J2000.00)} & 
\colhead{(pc)} & 
\colhead{(\kms{})}
}
\startdata
W3 IRS 4  & 02\tsp{h}25\tsp{m}30$\fs$92 & 62$^{\circ}$06$\arcmin$20$\farcs$70 & 1,950\tsp{1} & -35.0 \\
W3(OH) & 02 27 04.31 & 61 52 23.90 & 1,950\tsp{1} & -50.0 \\
L1448 IRS 3B & 03 25 36.33 & 30 45 15.00 & 232\tsp{2} & 8.0 \\
NGC 1333 IRAS 4A & 03 29 10.52 & 31 13 31.00 & 235\tsp{2} & 6.8 \\
NGC 1333 IRAS 4B & 03 29 12.00 & 31 13 07.80 & 235\tsp{2} & 6.8 \\
L1551 IRS 5 & 04 31 34.07 & 18 08 05.10 & 140\tsp{3} & 6.4 \\
Orion-KL & 05 35 14.46 & -05 22 27.50 & 418\tsp{4} & 9.0 \\
Orion-N & 05 35 14.45 & -05 21 03.20 & 418\tsp{4} & 9.0 \\
Orion-S & 05 35 13.55 & -05 24 08.30 & 418\tsp{4} & 6.5 \\
OMC-2 IRS 4 & 05 35 27.32 & -05 09 39.6 & 430\tsp{5} & 11.0 \\
IRAS 05338-0624 & 05 36 18.42 & -06 22 06.20 & 480\tsp{6} & 7.1 \\
NGC 2024 & 05 41 44.53 & -01 55 02.90 & 415\tsp{5} & 11.0 \\
NGC 2024 FIR 5 & 05 41 44.47 & -01 55 42.50 & 415\tsp{5} & 11.0 \\
NGC 2024 FIR 6 & 05 41 45.16 & -01 56 04.80 & 415\tsp{5} & 11.0 \\
NGC 2071 IR & 05 47 04.85 & 00 21 43.00 & 390\tsp{5} & 10.0 \\
S255N & 06 12 53.67 & 18 00 26.60 & 1,590\tsp{7} & 8.0 \\
G34.26+0.15 & 18 53 18.60 & 01 14 58.40 & 3,700\tsp{8} & 58.7 \\
S68N & 18 29 47.91 & 01 16 46.50 & 230\tsp{9} & 8.8 \\
W51M & 19 23 44.01 & 14 30 34.10 & 5,410\tsp{10} & 55.0 \\
DR 21(OH) & 20 39 01.09 & 42 22 48.90 & 2,000\tsp{11} & -5.0 \\
Cep A East & 22 56 18.15 & 62 01 46.00 & 700\tsp{1} & -10.1 \\
NGC 7538 IRS 1 & 23 13 45.34 & 61 28 10.50 & 2,650\tsp{1} & -57.3 \\
NGC 7538 IRS 9 & 23 14 01.66 & 61 27 20.00 & 2,650\tsp{1} & -57.3 \\
\enddata
\tablerefs{(1)\citealt{Reid09} (2)\citealt{Hir08} (3)\citealt{Ken94} (4)\citealt{Kim08} (5)\citealt{AT82} (6)\citealt{Chen93} (7)\citealt{Rygl10} (8)\citealt{KB94} (9)\citealt{Eir08} (10)\citealt{Sato10} (11)\citealt{OS93}}
\label{tab:sources}
\end{deluxetable*}

Details on the utility of \f{} as a high-density probe are given in \S\ref{h2coProbe}. 
In \S\ref{observations} our observational and calibration procedures are presented 
with a brief discussion of the measurement results in \S\ref{results}. \S\ref{analysis} 
describes the details of our analysis, including a discussion of the Large Velocity Gradient
(LVG) and Local Thermodynamic Equilibrium (LTE) approximations used, as 
well as source-by-source comparisons to past studies. A detailed discussion of the 
limitations of measuring density with the $J=3$ and 4 \textit{K}-doublet transitions 
of \f{} is provided in \S\ref{discussion}. 

\section{Formaldehyde as a High-Density Probe}
\label{h2coProbe}
	
Formaldehyde (\f{}) is a proven tracer of the high-density environs of 
molecular clouds. It is ubiquitous; \f{} is associated with 80\% of the HII 
regions surveyed by \citet{Down80} and possesses a large number of 
observationally accessible transitions from centimeter to far-infrared 
wavelengths. Because \f{} is a slightly asymmetric rotor molecule, most 
rotational energy levels are split by this asymmetry into two energy levels. 
Therefore, the energy levels must be designated by a total angular momentum 
quantum number, $J$, the projection of $J$ along the symmetry axis for a limiting 
prolate symmetric top, $K_{-1}$, and the projection of $J$ along the symmetry 
axis for a limiting oblate symmetric top, $K_{+1}$. This splitting leads to two 
basic types of transitions: the high-frequency $\Delta J = 1$, $\Delta K_{-1} 
= 0$, $\Delta K_{+1} = -1$ ``$P$-branch transitions'' and the lower frequency 
$\Delta J = 0$, $\Delta K_{-1} = 0$, $\Delta K_{+1} = \pm 1$ ``$Q$-branch'' 
transitions, popularly known as the ``$K$-doublet'' transitions (see discussion 
in \citealt{MW93}). The $P$-branch transitions are only seen in emission in regions 
where \textit{n}(H\tsb{2}) $\gtrsim$ 10\tsp{4} \cm{}. The excitation of the 
$K$-doublet transitions, however, is not so simple. For \textit{n}(H\tsb{2}) 
$\lesssim$ 10\tsp{5.5} \cm{}, the lower energy states of the \jone{} 
through \jfive{} \textit{K}-doublet transitions become 
overpopulated due to a collisional selection effect \citep{Evans75,Gar75}. 
This overpopulation cools the $J \leq 5$ $K$-doublets to excitation 
temperatures lower than that of the cosmic microwave background, causing 
them to appear in absorption. For \textit{n}(H\tsb{2}) $\gtrsim$ 10\tsp{5.5} 
\cm{}, this collisional pump is quenched and the $J \leq 5$ $K$-doublets 
are then seen in emission over a wide range of kinetic temperatures and 
abundances.

\section{Observations}
\label{observations}
	
The measurements reported here were made using the National Radio Astronomy 
Observatory (NRAO\footnote{The National Radio Astronomy Observatory is
  a facility of the National Science Foundation operated under
  cooperative agreement by Associated Universities, Inc.}) Green Bank
Telescope (GBT) during the periods 2007 January  
20-31, February 19, and 2008 July 2. Single pointing measurements were obtained 
of the \jthree (28.974805 GHz, $\theta_B=26\arcsec$) and \jfour (48.284547 GHz, 
$\theta_B=16\arcsec$) \textit{K}-doublet transitions of \f{} toward a sample of 
23 galactic star-forming regions (Table~\ref{tab:sources}). The 29 and 48 GHz measurements 
were made using the dual-beam Ka- (beam separation 78$\arcsec$) and Q-band (beam separation 58$\arcsec$) receivers, 
respectively, over 50 MHz of bandwidth sampled by 16,384 channels. The 
position-switching technique was employed with reference position located 30$\arcmin$ west in 
azimuth from each source position. The correlator configuration produced a spectral 
channel width of 3.052 kHz, which is approximately 0.03 and 0.02 \kms{} at 29 and 
48 GHz, respectively. The 9\tsb{2}--8\tsb{1} A\tsp{-} (28.969942 GHz) vibrational 
transition of CH\tsb{3}OH was also captured in the Ka-band spectra. 

To calibrate the intensity scale of our measurements, several corrections need to be 
considered:

\textit{Opacity.}--Historical opacity estimates based on atmospheric model calculations 
using ambient pressure, temperature, and relative humidity measurements indicated 
that $\tau{}_0$ at 29 and 48 GHz were, respectively, $\sim$ 0.05 and 0.25 during our 
observations. The respective opacity corrections exp[$\tau_0$csc(\textit{EL})] for the
range of source elevations (28--75$^{\circ}$) averaged 1.069 and 1.485, and were applied 
uniformly since the variance due to elevation was less than the absolute uncertainty in 
amplitude calibration (see below). 

\textit{Flux.}--Assuming point-source emission, one can use the current relation 
(derived from measurement) for the aperture efficiency 
\[\eta_A=0.71\exp(-[0.0163\nu(\mathrm{GHz})]^2)\] to convert antenna temperature to flux 
density. At 28.97 and 48.28 GHz this yields $\eta_A=0.568$ and 0.382, respectively. For 
elevation 90$^{\circ}$ and zero atmospheric opacity, $T_A/S=2.846\eta_A=1.61$ and 1.09
for  29 and 48 GHz, respectively. These are the degrees Kelvin per Jansky calibration
factors used to convert our spectra to flux assuming point-source emission. Incorporating 
atmospheric opacity and telescope efficiency, we have $T_A^*=T_A\exp{(A\tau_0)}/\eta_l=1.08T_A$
(29 GHz) and 1.50$T_A$ (48 GHz), where $\eta_l=0.99$ for the GBT. Finally, using 
$\eta_{mb} \simeq 1.32 \eta_A$, we can write the main beam brightness temperature as 
$T_{mb} \simeq T^*_A/\eta_{mb}=1.45T_A$ (29 GHz) and 2.90$T_A$ (48 GHz).

\textit{Source Structure.}--Since the beam sizes are about 63\% different, and we anticipate 
source sizes comparable to our beam sizes, it is necessary 
to scale both measurements to the same beam size to properly intercompare them. The 
beam coupling correction factor for a Gaussian source and a Gaussian beam to a 
measurement which has already been corrected to the main-beam efficiency scale 
is: \[ f_{\mathrm{couple}}=\frac{\theta_{\mathrm{source}}^2+\theta_{\mathrm{beam}}^2}{2\theta_{\mathrm{source}}^2} \]
Therefore, to scale the $J=3$ $T_{mb}$ measurements ($\theta_{\mathrm{beam}}=26\arcsec$) 
to a source size equal to the beam size at $J=4$ ($\theta_{\mathrm{beam}}=16\arcsec$), an 
additional factor of 1.82 must be applied to the $J=3$ $T_{mb}$ measurements. With this, we have assumed 
that each object in our sample spans 16$\arcsec$. Refer to 
\S\ref{SourceSize} for additional justification of this assumption.

\textit{Absolute amplitude calibration.}--The GBT absolute amplitude calibration is reported 
to be 10-15\% at all frequencies, limited mainly by temporal drifts in the noise diodes used 
as absolute amplitude calibration standards.

\LongTables
\begin{deluxetable*}{l c c c c c}
\tabletypesize{\scriptsize}
\tablecolumns{6} 
\tablewidth{0pc} 
\tablecaption{\f{} Measurement Results} 
\tablehead{ 
\colhead{} & 
\colhead{} & 
\colhead{$T_A^*$\tablenotemark{a}} & 
\colhead{\lsr{}} & 
\colhead{FWHM} & 
\colhead{log$\left[\frac{N(\small{\mathrm{ortho-\f{}}})}{\Delta\nu}\right]$\tablenotemark{b}}  
\\ 
\colhead{Source} & 
\colhead{Transition} & 
\colhead{(K)} & 
\colhead{(\kms{})} & 
\colhead{(\kms{})} & 
\colhead{[cm\tsp{-2} (\kms{})\tsp{-1}]}
}
%\startdata		
W3 IRS 4                 & \jfour{}    & $\leq$ 0.090 ($3\sigma$)  & \nodata                  & \nodata              & \nodata \\
				               & \jthree{}  & -0.023 $\pm$ 0.010 & -47.28 $\pm$ 0.44  & 6.66 $\pm$ 0.91 & 13.17\pms{0.16}{0.25} \\
				               &                 & -0.068 $\pm$ 0.010 & -42.53 $\pm$ 0.05  & 1.00 $\pm$ 0.12 & 13.65\pms{0.06}{0.07}\\
W3(OH)                  & \jfour{}    & 0.216 $\pm$ 0.028   & -47.56 $\pm$ 0.15  & 6.23 $\pm$ 0.15 & 14.18\pms{0.05}{0.06} \\
				               &                 & -0.136 $\pm$ 0.028 & -44.46 $\pm$ 0.15  & 1.46 $\pm$ 0.15 & 13.98\pms{0.08}{0.10}  \\
				               & \jthree{}  & 0.124 $\pm$ 0.014  & -47.60\tablenotemark{c}  & 6.20\tablenotemark{c} & 14.09$\pm0.05$ \\
				               &                 & -0.284 $\pm$ 0.014 & -44.50\tablenotemark{c}  & 1.50\tablenotemark{c} & 14.45$\pm0.02$ \\
				               &                 & -0.167 $\pm$ 0.014 & -46.96 $\pm$ 0.08  & 3.20 $\pm$ 0.34 & 14.22\pms{0.3}{0.4} \\				                              
L1448 IRS 3B          & \jfour{}    & $\leq$ 0.202 ($3\sigma$)   & \nodata                  & \nodata               & \nodata \\
				               & \jthree{}  & 0.041 $\pm$ 0.005  & 4.85 $\pm$ 0.06  & 1.76 $\pm$ 0.13 & 13.51\pms{0.5}{0.6} \\
NGC 1333 IRAS 4A  & \jfour{}    & 0.084 $\pm$ 0.031  & 7.00\tablenotemark{c}  & 2.06\tablenotemark{c} & 13.48\pms{0.14}{0.20} \\
				                &                & 0.039 $\pm$ 0.031  & 7.00\tablenotemark{c}  & 8.50\tablenotemark{c} & 13.15\pms{0.25}{0.67}  \\
	                            & \jthree{} & 0.162 $\pm$ 0.006  & 6.90\tablenotemark{c}  & 2.06 $\pm$ 0.06 & 13.85$\pm0.02$ \\
				                &                & 0.033 $\pm$ 0.006  & 6.90\tablenotemark{c}  & 8.50\tablenotemark{c} & 13.16\pms{0.07}{0.08} \\
	                            &                & -0.118 $\pm$ 0.006 & 7.50\tablenotemark{c}  & 2.00\tablenotemark{c} & 13.71$\pm0.02$ \\
NGC 1333 IRAS 4B   & \jfour{}   & 0.133 $\pm$ 0.029  & 7.08 $\pm$ 0.09  & 2.25 $\pm$ 0.26 & 13.83\pms{0.09}{0.11} \\
				                & \jthree{} & 0.036 $\pm$ 0.009  & 6.90\tablenotemark{c}  & 2.20\tablenotemark{c} & 13.40\pms{0.10}{0.13} \\
L1551 IRS 5              & \jfour{}   & 0.103 $\pm$ 0.039  & 6.51 $\pm$ 0.07  & 0.48 $\pm$ 0.14 & 13.82\pms{0.14}{0.21} \\
				                & \jthree{} & 0.050 $\pm$ 0.006  & 6.34 $\pm$ 0.03  & 0.73 $\pm$ 0.08 & 13.65$\pm0.05$ \\
Orion-KL                  & \jfour{}   & 1.748 $\pm$ 0.064  & 10.27 $\pm$ 0.02  & 1.72 $\pm$ 0.04 & 15.19$\pm0.02$ \\
				               &                 & 0.551 $\pm$ 0.064  & 6.71 $\pm$ 0.14  & 13.53 $\pm$ 0.46 & 15.12$\pm0.05$ \\
				               &                 & 0.938 $\pm$ 0.064  & 8.30\tablenotemark{c}  & 2.71 $\pm$ 0.17 & 14.92$\pm0.03$ \\
				               & \jthree{}  & 0.670 $\pm$ 0.016  & 10.41 $\pm$ 0.001  & 1.62 $\pm$ 0.003 & 14.93$\pm0.01$ \\
				               &                 & 0.563 $\pm$ 0.016  & 6.72 $\pm$ 0.03  & 9.58 $\pm$ 0.004 & 15.32$\pm0.01$ \\
				               &                 & 3.972 $\pm$ 0.016  & 8.16\tablenotemark{c}  & 2.46 $\pm$ 0.003 & 15.71$\pm0.01$ \\
Orion-N                   & \jthree{}  & 0.259 $\pm$ 0.012  & 9.37 $\pm$ 0.02  & 1.47 $\pm$ 0.05 & 13.95$\pm0.02$  \\
Orion-S                   & \jfour{}    & 0.793 $\pm$ 0.035  & 6.57\tablenotemark{c}  & 2.68 $\pm$ 0.09 & 14.70$\pm0.02$ \\
				               &                 & 0.243 $\pm$ 0.035 & 6.90\tablenotemark{c}  & 6.35 $\pm$ 0.40 & 14.19\pms{0.06}{0.07} \\
				               & \jthree{}  & 0.575 $\pm$ 0.011 & 6.39\tablenotemark{c} 0.00  & 2.65 $\pm$ 0.04 & 14.70$\pm0.01$ \\
				               &                 & 0.143 $\pm$ 0.011 & 6.72\tablenotemark{c}  & 6.35\tablenotemark{c} & 14.10\pms{0.03}{0.04} \\
OMC-2 IRS 4           & \jfour{}    & 0.175 $\pm$ 0.026  & 11.57 $\pm$ 0.05  & 1.66 $\pm$ 0.13 & 13.98\pms{0.06}{0.07} \\
				               & \jthree{} & 0.060 $\pm$ 0.008 & 11.50 $\pm$ 0.12  & 1.41 $\pm$ 0.05 & 13.65\pms{0.05}{0.06} \\
				               &                & 0.253 $\pm$ 0.008  & 11.36 $\pm$ 0.01  & 6.06 $\pm$ 0.49 & 14.27$\pm0.01$ \\				                             
IRAS 05338-0624    & \jfour{}   & 0.030 $\pm$ 0.029  & 9.60\tablenotemark{c}  & 2.48\tablenotemark{c} & 13.26\pms{0.29}{1.30} \\
				               &                & 0.075 $\pm$ 0.029  & 7.08\tablenotemark{c}  & 1.94\tablenotemark{c} & 13.65\pms{0.14}{0.21} \\
				               & \jthree{} & 0.019 $\pm$ 0.008  & 9.60\tablenotemark{c}  & 2.48\tablenotemark{c} & 13.21\pms{0.15}{0.24} \\
				               &                & 0.070 $\pm$ 0.008  & 7.08\tablenotemark{c}  & 1.94\tablenotemark{c}& 13.77$\pm0.05$  \\
NGC 2024               & \jfour{}   & 0.149 $\pm$ 0.038  & 11.50 $\pm$ 0.08  & 1.25 $\pm$ 0.19 & 13.95\pms{0.10}{0.13} \\
NGC 2024 FIR 5      & \jfour{}   & 0.202 $\pm$ 0.041  & 11.20\tablenotemark{c}  & 1.97 $\pm$ 0.19 & 14.34\pms{0.08}{0.10} \\
				               & \jthree{} & 0.224 $\pm$ 0.008  & 11.10\tablenotemark{c}  & 2.04 $\pm$ 0.05 & 14.55$\pm0.01$ \\
				               &                & -0.062 $\pm$ 0.008 & 9.10\tablenotemark{c}  & 0.58 $\pm$ 0.06 & 13.99$\pm0.06$ \\
NGC 2024 FIR 6      & \jfour{}   & 0.108 $\pm$ 0.030  & 11.12 $\pm$ 0.12  & 2.56 $\pm$ 0.27 & 13.55\pms{0.11}{0.14} \\
				               & \jthree{} & 0.208 $\pm$ 0.011  & 11.10\tablenotemark{c}  & 2.14 $\pm$ 0.07 & 13.89$\pm0.02$ \\
				               &                & -0.044 $\pm$ 0.011 & 9.11 $\pm$ 0.12  & 0.57 $\pm$ 0.11 & 13.21\pms{0.10}{0.13} \\
NGC 2071 IR           & \jfour{}   & 0.097 $\pm$ 0.040  & 8.33\tablenotemark{c}  & 4.79 $\pm$ 1.00 & 13.69\pms{0.15}{0.23} \\
				               &                & 0.155 $\pm$ 0.040  & 9.81 $\pm$ 0.13  & 1.70 $\pm$ 0.39 & 13.90\pms{0.10}{0.13} \\
				               & \jthree{} & 0.085 $\pm$ 0.008  & 8.33 $\pm$ 0.18  & 4.73 $\pm$ 0.27 & 13.77$\pm0.04$ \\
				               &                & 0.175 $\pm$ 0.008  & 9.58 $\pm$ 0.03  & 1.84 $\pm$ 0.12 & 14.08$\pm0.02$ \\
				               &                & -0.043 $\pm$ 0.008 & 16.17 $\pm$ 0.09  & 1.94 $\pm$ 0.26 & 13.48\pms{0.07}{0.09} \\
S255N                     & \jfour{}   & 0.181 $\pm$ 0.019  & 8.87 $\pm$ 0.05  & 2.94 $\pm$ 0.12 & 13.92\pms{0.04}{0.05} \\
				               & \jthree{} & 0.192 $\pm$ 0.008  & 8.79 $\pm$ 0.02  & 2.92 $\pm$ 0.06 & 14.06$\pm0.02$ \\
G34.26+0.15          & \jfour{}   & 0.147 $\pm$ 0.053  & 57.72 $\pm$ 0.24  & 5.15 $\pm$ 0.63 & 14.20\pms{0.13}{0.19} \\
S68N                      & \jfour{}   & 0.119 $\pm$ 0.040  & 7.85 $\pm$ 0.19  & 4.18 $\pm$ 0.42 & 13.58\pms{0.13}{0.18} \\
W51M                     & \jfour{}   & 0.251 $\pm$ 0.080  & 56.44 $\pm$ 0.25  & 8.28 $\pm$ 0.55 & 14.40\pms{0.12}{0.17} \\
DR 21(OH)              & \jfour{}   & 0.320 $\pm$ 0.034 & -4.22 $\pm$ 0.06  & 3.02 $\pm$ 0.14 & 14.54\pms{0.04}{0.05} \\
				               &                & 0.345 $\pm$ 0.034 & -1.28\tablenotemark{c}  & 3.40\tablenotemark{c} & 14.04\pms{0.04}{0.05} \\
				               & \jthree{} & 0.247 $\pm$ 0.005 & -4.45 $\pm$ 0.04  & 2.89 $\pm$ 0.06 & 14.59$\pm0.01$ \\
				               &                & 0.182 $\pm$ 0.005 & -1.28 $\pm$ 0.06  & 3.40 $\pm$ 0.11 & 13.76$\pm0.01$ \\
Cep A East             & \jfour{}   & 0.078 $\pm$ 0.033 & -10.40\tablenotemark{c}  & 2.40\tablenotemark{c} & 13.50\pms{0.15}{0.24} \\
				               &                & 0.042 $\pm$ 0.033 & -10.28 $\pm$ 0.59  & 6.00\tablenotemark{c} & 13.23\pms{0.25}{0.66} \\
				               & \jthree{} & 0.036 $\pm$ 0.007 & -10.43 $\pm$ 0.10  & 2.43 $\pm$ 0.23 & 13.27\pms{0.07}{0.09} \\
NGC 7538 IRS 1      & \jfour{}   & 0.111 $\pm$ 0.028  & -57.70 $\pm$ 0.14  & 4.18 $\pm$ 0.36 & 14.25\pms{0.10}{0.13} \\
				               & \jthree{} & -0.102 $\pm$ 0.009& -59.00\tablenotemark{c}  & 2.19 $\pm$ 0.13 & 14.39$\pm0.04$ \\
				               &                & 0.110 $\pm$ 0.009 & -57.65\tablenotemark{c}  & 4.00\tablenotemark{c} & 14.43\pms{0.03}{0.04} \\
NGC 7538 IRS 9      & \jfour{}   & $\leq$ 0.133 ($3\sigma$)  & \nodata                  & \nodata             & \nodata \\
				               &                & 0.048 $\pm$ 0.009 & -56.82 $\pm$ 0.12  & 3.31 $\pm$ 0.29 & 13.39\pms{0.07}{0.09} \\
\enddata
\tablenotetext{a}{Source radiation temperature corrected for telescope efficiency and atmospheric attenuation. Main beam efficiency and beam dilution calibrations applied for analysis (see \S\ref{observations}).}
\tablenotetext{b}{\textrm{L}TE approximation. See \S\ref{LTE} for details.}
\tablenotetext{c}{Quantity fixed in Gaussian fit.}
\label{tab:h2coMeas}
\end{deluxetable*}
		
\section{Results}
\label{results}
 
Both transitions were detected in 15 of the 23 sources in our sample. Three objects yielded nondetections
in the $J=4$ transition (W3 IRS 4, L1448 IRS 3B, and NGC 7538 IRS 9) and another five sources 
were observed in only one of the two transitions due to time 
constraints. Data reduction was accomplished using the CLASS (Continuum and Line Analysis
Single-dish Software) package from GILDAS\footnote{http://www.iram.fr/IRAMFR/GILDAS/}. 
The following spectra have been smoothed to 0.2526 \kms{} (\jthree{}) and 0.1516 \kms{} 
(\jfour{}) to increase the signal-to-noise ratio of the individual channels. Each line profile was 
fitted with 1--3 Gaussian components, and each velocity component was checked for 
consistency with previous measurements (see \S\ref{comparisons}). 

\begin{figure*}
 \figurenum{1}
 \epsscale{1}
 \plotone{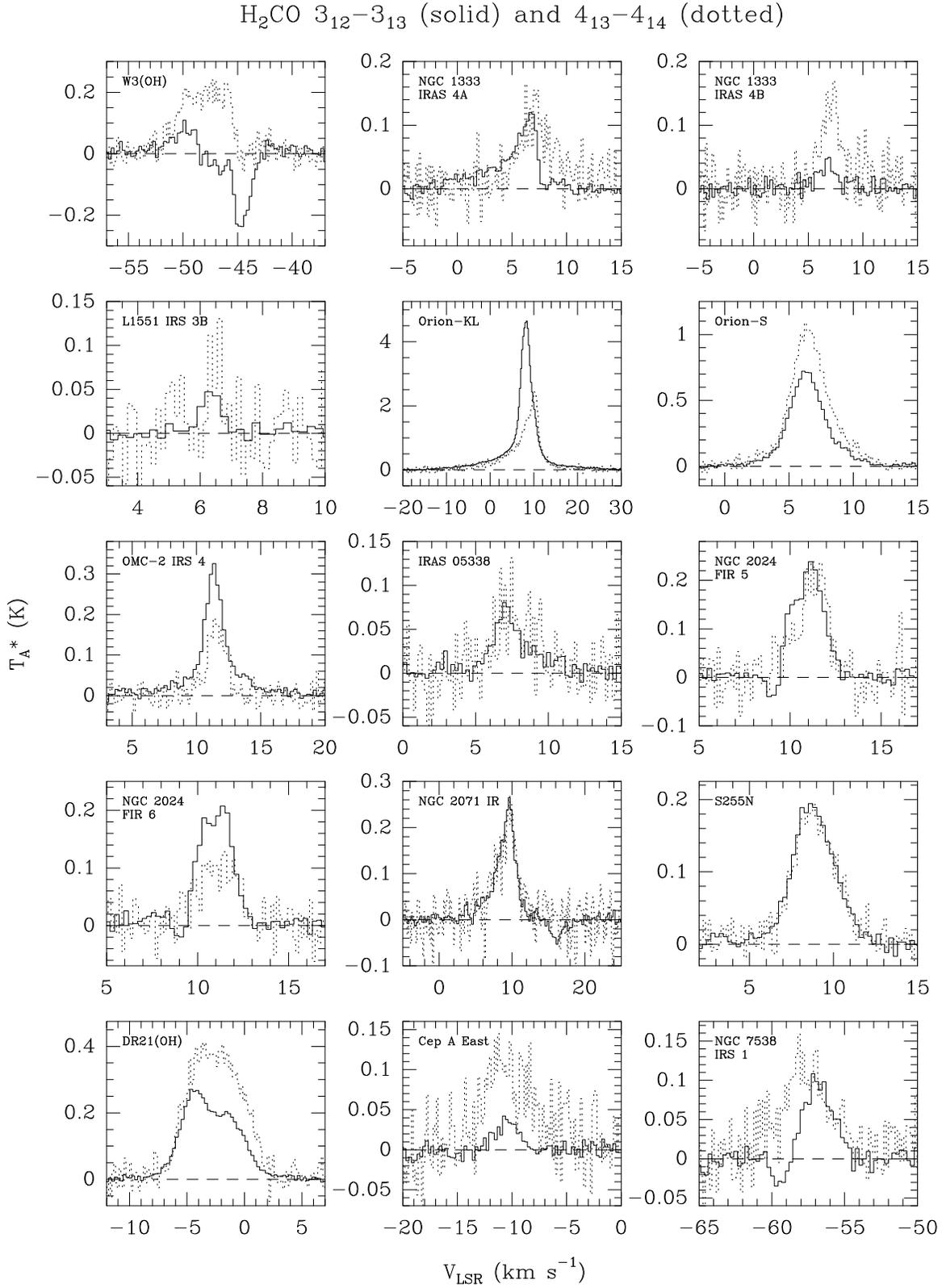}
 \caption{Spectra for sources toward which both transitions were detected.}
 \label{fig:DualDetections}
 \end{figure*}

\begin{figure}
 \figurenum{2}
 \epsscale{1}
 \plotone{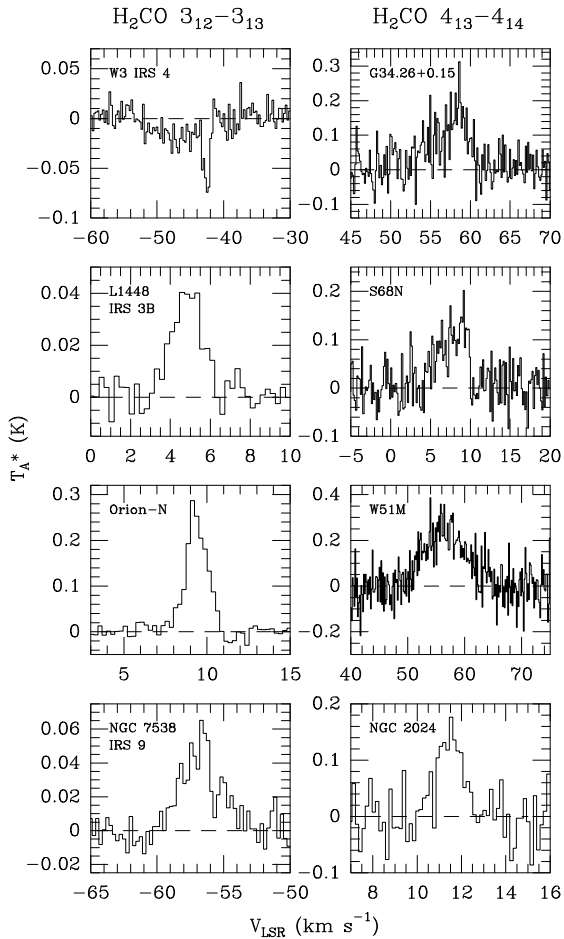}
 \caption{Spectra for sources toward which only one transition was detected or observed.}
 \label{fig:SingleDetections}
 \end{figure}

Table~\ref{tab:h2coMeas} lists the peak intensity ($T_A^*$), velocity of the 
local standard of rest (\lsr{}), and velocity width at the half maximum (FWHM) determined 
via Gaussian fitting for each object. LTE column 
densities are also listed (see \S\ref{LTE}). Note that the peak intensities 
quoted in Table~\ref{tab:h2coMeas} are given in the $T_A^*$ scale in the event 
that a reader may wish to apply calibration factors different than those used in our analysis 
(see \S\ref{observations}). The spectra for sources detected in both the \jthree{} and \jfour{} 
transitions are displayed in Figure~\ref{fig:DualDetections}, while spectra from sources detected or observed in only 
one of the two transitions are shown in Figure~\ref{fig:SingleDetections}. 

\begin{figure}
 \figurenum{3}
 \epsscale{1}
 \plotone{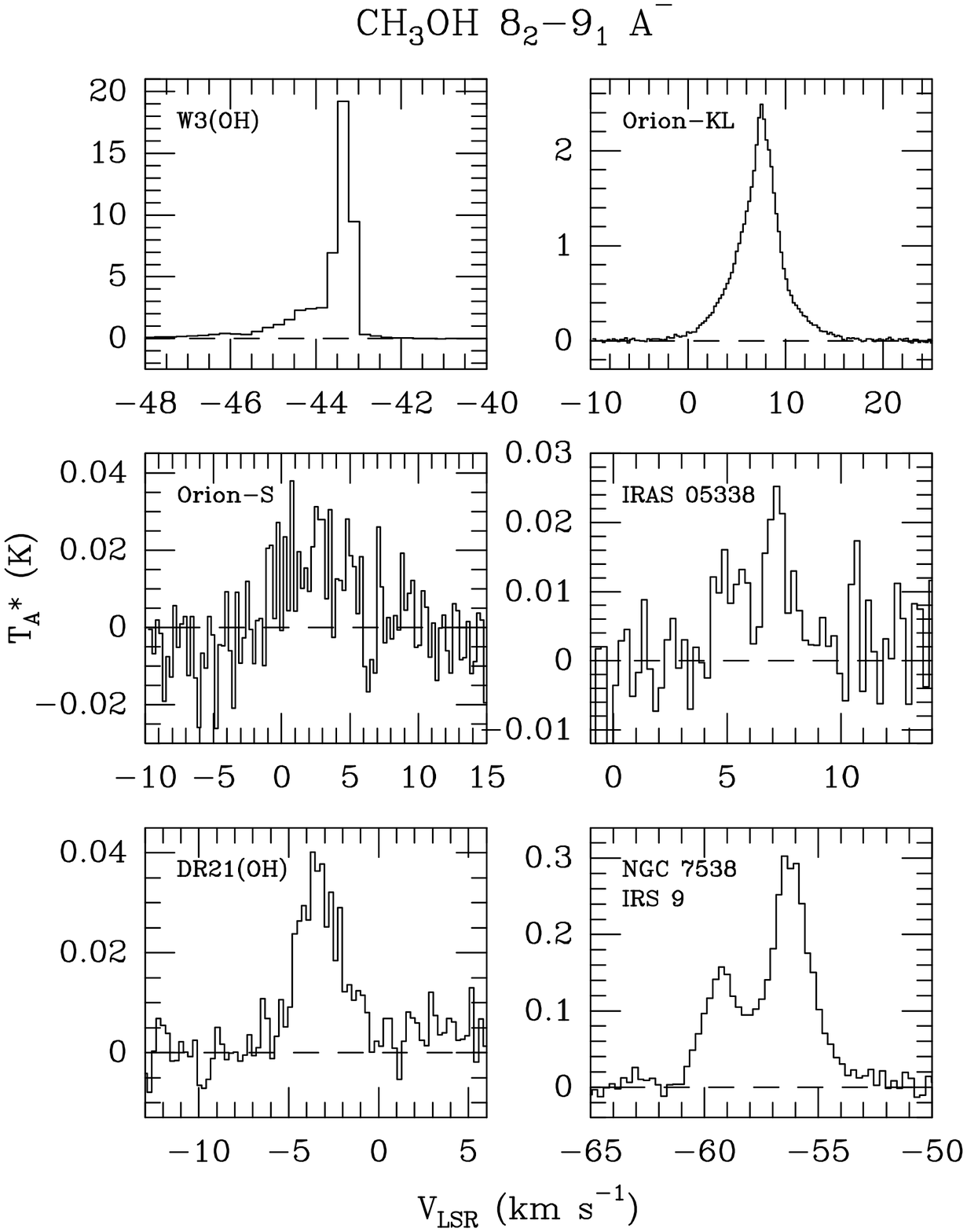}
 \caption{Methanol spectra.}
 \label{fig:ch3oh}
 \end{figure}

The 9\tsb{2}--8\tsb{1} A\tsp{-} transition of 
CH\tsb{3}OH was detected in 6 sources. Since no further analysis is presented, Gaussian 
components were not fit to each spectrum and the basic line
parameters, determined via direct measurement of the line peak and
width, are listed in 
Table~\ref{tab:ch3ohMeas} with the spectra displayed in Figure~\ref{fig:ch3oh}. Maser emission of this transition was detected toward 
one source, W3(OH). 

% \onecolumn
% \twocolumn
 
\section{Analysis}
\label{analysis}

\subsection{Spatial and Column Density Derivation from LVG Modelling}
\label{LVG}
	
To derive the spatial density [H\tsb{2} number density, \textit{n}(H\tsb{2})] 
and ortho-\f{} column density in our sample of galactic star-forming regions, we 
have used a model which incorporates the Large Velocity Gradient (LVG) 
approximation \citep{Sob60} to radiative transfer in molecular clouds. The detailed properties of 
our implementation of the LVG approximation are described in \citet{MW93}. In short, the model is used 
to predict the individual brightness temperatures of both the $J=3$ and 4 transitions as well as the ``transition ratio,''
R$_i=\int{}T_{mb}(\jthree)d\nu$/$\int{}T_{mb}(\jfour)d\nu$,
for a given set of physical conditions: spatial density, column density, kinetic temperature, and optical depth. 
The solution is then indicated by the intersection of the brightness temperature and R$_i$ predictions. 
This is illustrated by Figure~\ref{fig:LVG}, which is discussed in more detail at the end of this section. It is important 
to note the behavior of the transition ratio. Since we calculate R$_i$ as (lower excitation)/(higher excitation), a higher ratio implies 
lower spatial density and/or kinetic temperature. 
As noted by 
\citet{MW93}, one of the major sources of uncertainty in an LVG model 
prediction of physical conditions is the uncertainty associated with the 
collisional excitation rates used. 

Our model incorporates the scaled excitation 
rates involving \f{} and He calculated by \citet{Gre91}. An important point 
regarding our implementation of the LVG model is the scaling of the calculated 
\f{}/He excitation rates to those appropriate for collisions with H\tsb{2}. Following 
the recommendation of \citet{Gre91}, we scale the calculated He rates by a factor 
of 2.2 to account for (1) the reduced collision velocity of He relative to H\tsb{2}, 
which scales as the inverse-square-root of the masses of He and H\tsb{2}, and (2) 
the larger cross section of the H\tsb{2} molecule ($\sim1.6$; \citealt{Nerf75}) 
relative to He. With these scaling factors, \citet{Gre91} suggests that the total 
collisional excitation rate for a given \f{} transition is accurate to $\sim20\%$. 
Thus the physical conditions predicted by our LVG model are limited to an accuracy 
of no better than 20\%. It should be noted that the 
collision rates of \f{} with H\tsb{2} have been rederived with a claimed accuracy of 
10\% by \citet{Tro09}, though only for temperatures $<30$ K. 

It is also important to note that the LVG model for the \jthree{} and \jfour{} transitions of \f{} is somewhat 
dependent on kinetic temperature (\textit{T}\tsb{K}). As such, it was important 
in this study to carefully select temperature estimates from the literature that 
best reflect the material being traced by our observations. The temperatures used 
can be found with their object references in \S\ref{comparisons}. The \textit{T}\tsb{K} dependence of the 
LVG model and its effect on our measurements is explored in detail in \S\ref{temp}.

LVG results for the 15 sources toward which both transitions were detected are 
displayed in Table~\ref{tab:LVGDetections}. A range of derived densities based on the 
range in our adopted kinetic temperatures is provided in column 5 to illustrate the kinetic temperature 
dependence of the LVG model. In absence of error bars on the \textit{T}\tsb{K} values assumed from the literature, a range of $\pm$ 50\% was used. 
Column 6 includes the range of density estimates found in selected literature (see \S\ref{comparisons}).
Note that the uncertainties associated with our results reflect only 
measurement error. Additional uncertainties associated with our kinetic temperature and spatial 
extent assumptions are described in \S\ref{discussion}.

For 6 of 21 velocity components, the observed transition ratio [R$_i=\int{}T_{mb}(\jthree)d\nu$/$\int{}T_{mb}(\jfour)d\nu$] and brightness 
temperatures ($T_{mb}$) were not consistent with LVG 
predictions and a Local Thermodynamic Equilibrium (LTE) approximation 
was employed to place a limit on the density (see \S\ref{LTE}). In general, this occurred when 
the R$_i$ was low ($<1$), meaning that the higher excitation $J=4$ transition 
was unexpectedly observed to be significantly brighter than the $J=3$. 
Figure~\ref{fig:LVG} displays $\mathrm{|LVG~Prediction - Observation|}$ for the transition ratio 
and $J=4$ brightness temperature in units of the measurement error $\sigma$ for two sources. OMC-2 IRS 4 (left panel) 
is indicative of those sources with observed R$_i$ $\sim$ 1--2 for which the density could be 
fully constrained. The solution is indicated by the intersection of the violet and 
shaded ($<1\sigma$) regions. Cep A East (right panel) is representative of those objects with ratios $\lesssim1$. 
Note how the color and solid-line contours are parallel. Further discussion of the reasons 
 for these failed model fits is given in \S\ref{discussion}.  
 
 \begin{deluxetable*}{l r r r r}	
\tabletypesize{\scriptsize}
\tablecolumns{5}	
\tablewidth{0pc}
\tablecaption{CH\tsb{3}OH 9\tsb{2}--8\tsb{1} A\tsp{-} Measurement Results}
\tablehead{ 
\colhead{} & 
\colhead{$T_A^*$} & 
\colhead{\lsr{}} & 
\colhead{FWZI} & 
\colhead{$\int{}T_A^*d\nu$} 
\\ 
\colhead{Source} & 
\colhead{(K)} & 
\colhead{(\kms{})} & 
\colhead{(\kms{})} & 
\colhead{(K \kms{})}
}
\startdata
W3(OH)\tablenotemark{a} & 19.211 $\pm$ 0.014 & -43.40 $\pm$ 0.01 & 3.06 $\pm$ 0.01 & 12.093 $\pm$ 0.012 \\
Orion-KL\tablenotemark{a} & 2.483 $\pm$ 0.016 & 7.53 $\pm$ 0.01 & 18.18 $\pm$ 0.04 & 11.089 $\pm$ 0.034 \\
Orion-S & 0.025 $\pm$ 0.011 & 2.39 $\pm$ 0.40 & 12.63 $\pm$ 0.82 & 0.104 $\pm$ 0.020 \\
IRAS 05338-0624 & 0.021 $\pm$ 0.008 & 7.19 $\pm$ 0.36 & 9.73 $\pm$ 0.91 & 0.061 $\pm$ 0.013 \\
DR21(OH) & 0.036 $\pm$ 0.005 & -3.28 $\pm$ 0.09 & 9.22 $\pm$ 0.23 & 0.115 $\pm$ 0.008 \\
NGC 7538 IRS 9\tablenotemark{a} & 0.292 $\pm$ 0.009 & -56.20 $\pm$ 0.02 & 8.82 $\pm$ 0.06 & 0.977 $\pm$ 0.013 \\
\enddata	
\tablenotetext{a}{Multiple velocity components present. See Figure~\ref{fig:ch3oh} for spectra.}
\label{tab:ch3ohMeas}
\end{deluxetable*}

\subsection{LTE Approximation}
\label{LTE}

To supplement the LVG model and provide a useful check on its results, a Local 
Thermodynamic Equilibrium (LTE) approximation to the column density, 
$N(\mathrm{ortho-\f{}})/\Delta\nu$ cm\tsp{-2} (\kms{})\tsp{-1}, 
has been calculated for each velocity component detected (listed in Table~\ref{tab:h2coMeas}):
\[ N=\frac{3k}{8\pi\nu{}S\mu^2}\frac{Q_{rot}}{g_{u}g_{K}g_{nuclear}}\exp\left(\frac{E_u}{kT_{ex}}\right)\times T_{mb} \]
\[ Q_{rot}=\frac{1}{3}\left(\frac{\pi k^3}{h^3 AB^2}\right)^{1/2}T^{3/2}, hA\ll kT \]

\noindent{}where the line strength S = $\frac{1}{12}$ ($J=3$), $\frac{1}{20}$ ($J=4$); 
dipole moment $\mu$ = 2.331 debye; rotational degeneracy 
$g_{u}$ = 7 ($J=3$), 9 ($J=4$); $K$ degeneracy $g_{K}$ = 2 (for $K\neq$ 0 in 
symmetric top molecules); nuclear spin degeneracy $g_{nuclear}$ = 3 
(for ortho-\f{}); level energy above ground \textit{E}\tsb{u} = 33.479 K ($J=3$), 
47.928 K ($J=4$). Note that the peak intensity ($T_{mb}$) is 
used instead of integrated intensity to give column density \textit{per unit line width} 
for consistency with the LVG model results. Note also that the LTE approximation assumes 
optically thin emission with excitation temperature equivalent to kinetic temperature. 

The former assumption 
is reasonable given that \f{} transitions rarely become optically thick due principally to fairly low abundances. This property, in part, 
makes \f{} an ideal probe of the dense regions within molecular clouds (see \S\ref{h2coProbe}). Given the average  physical conditions 
derived for our sample [\textit{n}{(H\tsb{2})} $\sim$ 10\tsp{6} \cm{}, $N(\small{\mathrm{ortho-\f{}}})/\Delta\nu\sim10^{14}$ cm\tsp{-2} (\kms{})\tsp{-1}, $T_K=100$ K], 
the LVG-predicted optical depth ($\tau$) $\sim$ 0.078. The maximum value found for our sample is 0.27, and transitions can generally be considered optically thin for $\tau<0.4$. 

The second assumption that \textit{T}\tsb{ex} = \textit{T}\tsb{K} is
more troublesome. For \textit{n}{(H\tsb{2})} $\sim$ 10\tsp{6} \cm{}
and $N(\small{\mathrm{ortho-\f{}}})/\Delta\nu\sim10^{14}$ cm\tsp{-2}
(\kms{})\tsp{-1}, the excitation temperatures for the $J=3$ and 4
\textit{K}-doublets are on order $\frac{T_K}{10}$\,K, while
\textit{T}\tsb{ex}$\sim\frac{T_K}{2}$\,K for the $\Delta J = 1$, $\Delta K_{-1} 
= 0$, $\Delta K_{+1} = -1$ $P$-branch transitions. Therefore, since
the ortho-H$_2$CO column density is mainly dependent upon the
excitation temperature of the $P$-branch transitions, \textit{T}\tsb{ex} 
$\lesssim$ \textit{T}\tsb{K} for our sample and the resulting LTE column
densities must then, strictly speaking, be considered upper limits
given that \textit{T}\tsb{mb} $\propto$ \textit{T}\tsb{ex}(1 -
exp(-$\tau$)). Nevertheless, the LTE-predicted column densities are
generally within a factor of 2 of the LVG results  
(see \S\ref{SingleLine} for more details). We consider the LTE approximation to be a reasonable one in cases where the LVG model failed to 
fully constrain the physical conditions. 
See \S\ref{comparisons} for kinetic temperature assumptions (alternatively, Tables~\ref{tab:LVGDetections}--\ref{tab:LVGSingleLine}).

For those cases (described in the previous section) in which LVG modeling was unsuccessful, 
the unique properties of the \textit{K}-doublet transitions discussed in 
\S\ref{h2coProbe} were used to place a limit on the spatial density. A column density was assumed 
using the LTE approximation described above, and the LVG model was then employed to determine the 
minimum density required for both transitions to be observed in emission. 

The LTE approximation was used in a similar manner for the 12 velocity components 
detected in only one of the two observed transitions. With the exception of Cep A East, 
these components were detected in the \jthree{} transition and not the \jfour{}. 
If seen in absorption at $J=3$, the undetected $J=4$ counterpart must also appear in absorption 
because of the excitation of the \f{} molecule. If observed in emission at $J=3$, the $J=4$ 
transition may appear in either state because the \jfour{} transition is subject to absorption 
beginning at slightly higher densities (see \S\ref{exeffects} for further detail). 
The range of densities permitting the $J=3$ transition to be in emission with 
the $J=4$ in absorption is very small, typically spanning 0.2 on the log scale. Rather 
than dictate this range for each object, we have opted to list just the limit corresponding 
to the $J=4$ emission/absorption boundary. Table~\ref{tab:LVGNondetections} displays 
these results. 

\subsection{Singe Transition \f{} Column Density Estimates}
\label{SingleLine}
		
Observations of only one transition were performed toward 5 sources. For these 
objects, the LVG model can be used to estimate the ortho-\f{} column density 
by assuming a kinetic temperature and spatial density. Kinetic temperatures 
were culled from the literature and can be found with their object references in \S\ref{comparisons}. 
A spatial density of 10\tsp{6.15} \cm{} 
was assumed based on an average of the results from those sources for 
which a full LVG analysis could be made. These results are presented in conjunction 
with the LTE approximation for comparison in Table~\ref{tab:LVGSingleLine}. 

%\onecolumn
 \begin{figure*}
 \figurenum{4}
 \epsscale{1.02}
 %Scaled to strange # because of vertical bars appeared across the image at \epsscale{1}
 \plotone{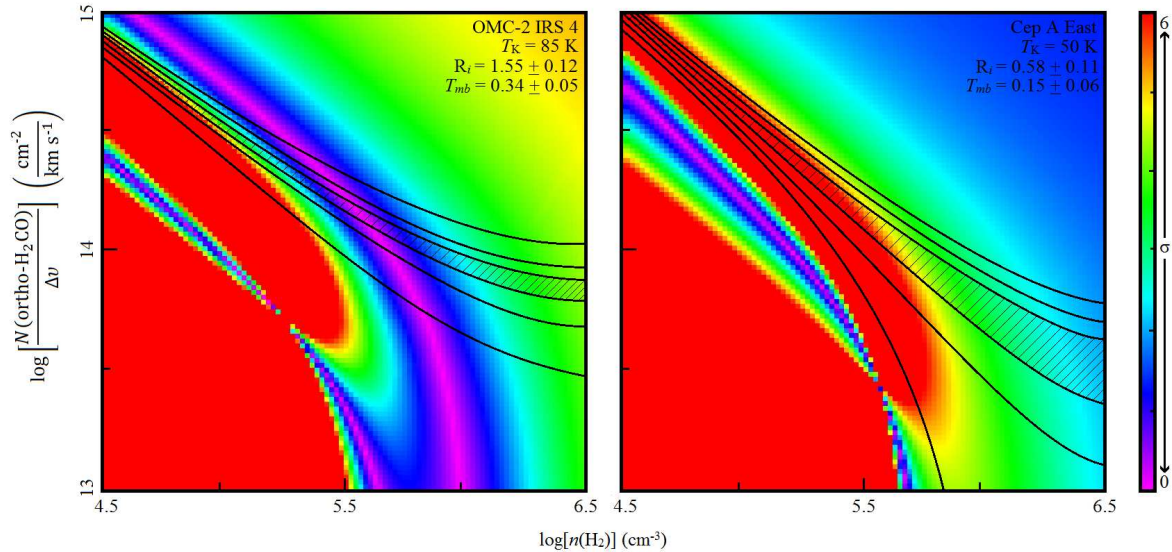}
 \caption{Comparison of model results for representative examples of the 15 velocity components with
 transition ratios (R$_i$) $\sim$ 1--2 that were well-fit (left panel) and the remaining 6 components with ratios $\lesssim1$ 
 whose observed profiles were inconsistent with LVG predictions (right panel). 
 \textit{Color}: $\mathrm{|LVG~Prediction - Observation}|$ for the integrated intensity ratio R$_i$ (\jthree{}/\jfour{}) in units 
 of the measurement error $\sigma$ from $<1\sigma$ (violet) to $>6\sigma$ (red). \textit{Contours}:
 $\mathrm{|LVG~Prediction - Observation|}$ for the \jfour{} brightness temperature $T_{mb}$ in units of $\sigma$; shaded region 
 corresponds to $\leq1\sigma$. Observed values listed in the upper right of each panel. Model solution indicated by the intersection of the 
 violet and shaded regions.}
 \label{fig:LVG}
 \end{figure*}
% \twocolumn

\begin{deluxetable*}{l c c c c c c c}

\tablecolumns{6}
\tabletypesize{\scriptsize}
\tablewidth{0pc}
\tablecaption{LVG Results: Velocity Components Detected in Both Transitions} 
\tablehead{ 
\colhead{} & 
\colhead{} & 
\colhead{} & 
\multicolumn{3}{c}{log[\textit{n}(H\tsb{2})]} & 
\multicolumn{2}{c}{log$\left[\frac{N(\small{\mathrm{ortho-\f{}}})}{\Delta\nu}\right]$} 
\\ 
\colhead{} & 
\colhead{\underline{$\int{}T_{mb}(\jthree)d\nu$}} & 
\colhead{\textit{T}\tsb{K}\tablenotemark{a}} & 
\multicolumn{3}{c}{(\cm{})} & 
\multicolumn{2}{c}{[cm\tsp{-2} (\kms{})\tsp{-1}]} 
\\ 
\colhead{Source} & 
\colhead{$\int{}T_{mb}(\jfour)d\nu$} & 
\colhead{(K)} & 
\colhead{\textit{T}\tsb{Best}} & 
\colhead{\textit{T}\tsb{Range}} & 
\colhead{Literature\tablenotemark{b}} &
\colhead{\jthree{}} & 
\colhead{\jfour{}}
}
\startdata		
W3(OH)$_{\mathrm{emission}}^{\mathrm{c}}$ & 0.73$\pm{0.03} $ & 110\pms{5}{15} & $\geq$4.92\pms{0.02}{0.06} & $\geq$4.88--5.00 & 5.95 & 14.09$\pm$0.05 & 14.18\pms{0.05}{0.06} \\
W3(OH)\tsb{absorption} & 2.70$\pm{0.33}$ & 60$\pm{20}$ & 4.11$\pm{0.05}$ & 3.97--4.21 & 4.70 & 14.64\pms{0.09}{0.18} & 14.64$\pm{0.07}$ \\
NGC 1333 IRAS 4A\tablenotemark{d} & 1.53$\pm{0.15}$ & 50 & 6.20\pms{0.18}{0.11} & 5.87--7.07 & 5.48--6.30 & 13.73\pms{0.03}{0.01} & 13.89\pms{0.08}{0.07} \\
NGC 1333 IRAS 4B\tablenotemark{c} & 0.34$\pm{0.06}$ & 80 & $\geq$5.31\pms{0.09}{0.05} & $\geq$5.10--5.62 & 5.48--6.30 & 13.40\pms{0.10}{0.13} & 13.83\pms{0.09}{0.11} \\
L1551 IRS 5 & 0.93$\pm{0.33}$ & 100\pms{50}{60} & 6.87\pms{>1.13}{0.76} & 6.49--$>$8.00 & 5.78--7.04 & 13.47\pms{0.03}{0.07} & 13.64\pms{0.14}{0.21} \\
Orion-KL\tsb{hot core} & 0.92$\pm{0.02}$ & 300\tsb{-70} & 5.97\pms{0.10}{0.09} & 6.08 & 5.70--6.95 & 14.57\pms{0.03}{0.01} & 14.42\pms{0.05}{0.08} \\
Orion-KL$\mathrm{_{compact~ridge}^c}$ & 4.85$\pm{0.12}$ & 135\pms{25}{15} & \nodata & \nodata & 5.50--5.70 & 15.71$\pm{0.01}$ & 14.93$\pm{0.03}$ \\
Orion-KL$\mathrm{_{extended~ridge}^c}$ & 0.45$\pm{0.01}$ & 135\pms{25}{15} & \nodata & \nodata & $>5.00$ & 14.93$\pm{0.01}$ & 15.19$\pm{0.02}$ \\
Orion-S\tsb{peak} & 0.91$\pm{0.02}$ & 100 & 7.21\pms{0.79}{0.40} & 6.43--$>$8.00 & 6.60--7.78 & 14.61$\pm{0.01}$ & 14.61\pms{0.04}{0.01} \\
Orion-S$_{\mathrm{wing}}^{\mathrm{c}}$ & 0.74$\pm{0.03}$ & 100 & $\geq$4.96\pms{0.02}{0.04} & $\geq$4.66--5.26 & \nodata & 14.10\pms{0.03}{0.04} & 14.19\pms{0.06}{0.07} \\
OMC-2 IRS 4 & 1.55$\pm{0.12}$ & 85\pms{25}{10} & 5.51$\pm{0.09}$ & 5.37--5.59 & 5.65--6.28 & 14.23$\pm{0.01}$ & 14.19\pms{0.03}{0.05} \\
IRAS 05338-0624\tablenotemark{d} & 1.05$\pm{0.16}$ & 95$\pm$55 & 6.49\pms{>1.51}{0.30} & 6.20--$>$8.00 & 5.30--6.00 & 13.59$\pm{0.05}$ & 13.51\pms{0.14}{0.19} \\
NGC 2024 FIR 5 & 1.45$\pm{0.13}$ & 160\pms{90}{60} & 5.41\pms{0.11}{0.12} & 5.31--5.57 & 5.95--6.00 & 14.13$\pm{0.01}$ & 14.15\pms{0.05}{0.07} \\
NGC 2024 FIR 6 & 2.03$\pm{0.23}$ & 40($<$160) & 5.83\pms{0.11}{0.07} & 4.94 & 5.95--6.30 & 14.07\pms{0.01}{0.03} & 14.03\pms{0.05}{0.09} \\
NGC 2071 IR\tsb{peak} & 1.54$\pm{0.19}$ & 80\pms{20}{15} & 5.71\pms{0.18}{0.14} & 5.57--5.85 & 5.48--6.48 & 14.03$\pm{0.01}$ & 14.05\pms{0.08}{0.09} \\
NGC 2071 IR\tsb{wing} & 1.09$\pm{0.13}$ & 80\pms{20}{15} & 6.53\pms{0.52}{0.26} & 6.34--6.73 & $>5.00$ & 13.59\pms{0.03}{0.05} & 13.57\pms{0.15}{0.22} \\
S255N & 1.33$\pm{0.06}$ & 70$\pm$30 & 6.02\pms{0.09}{0.05} & 5.78--6.58 & 6.48--7.20 & 13.97\pms{0.02}{0.01} & 13.97\pms{0.04}{0.03} \\
DR 21(OH)\tsb{MM1} & 0.93$\pm{0.04}$ & 160\pms{140}{90} & 6.34\pms{0.16}{0.13} & 6.04--$>$8.00 & 5.80--6.10 & 14.17$\pm{0.01}$ & 14.16\pms{0.05}{0.06} \\
DR 21(OH)$_{\mathrm{MM2}}^{\mathrm{c}}$ & 0.67$\pm{0.02}$ & 30($<$60) & $\geq$5.58$\pm0.08$ & $\geq$5.36 & 7.00--7.83 & 13.76$\pm$0.01 & 14.04\pms{0.04}{0.05} \\
Cep A East\tablenotemark{c} & 0.58$\pm{0.11}$ & 70\pms{30}{30} & $\geq$5.42\pms{0.02}{0.04} & $\geq$5.30--5.91 & 6.30-6.78 & 13.27\pms{0.07}{0.09} & 13.50\pms{0.15}{0.24} \\
NGC 7538 IRS 1 & 1.20$\pm{0.10}$ & 220\pms{30}{40} & 5.78\pms{0.15}{0.12} & 5.75--5.83 & 5.30--7.00 & 13.75\pms{0.03}{0.05} & 13.73\pms{0.09}{0.12} \\		
\enddata
\tablenotetext{a}{Culled from the literature; in absence of error estimates, a range of $\pm$50\% was tested. See \S\ref{comparisons} for references.}
\tablenotetext{b}{Spatial density range found in selected literature. See \S\ref{comparisons} for references.}
\tablenotetext{c}{Full LVG modeling was unsuccessful; LTE approximated column density quoted. See \S\ref{LTE} for details on limit derivation. }
\tablenotetext{d}{Total integrated emission used; see \S\ref{n1333} \& \S\ref{i05338} for details.}
\label{tab:LVGDetections}
\end{deluxetable*}
	
\subsection{Comparison to Previous Measurements}
\label{comparisons}
	
What follows are very brief descriptions of past studies of the objects in our sample that were most 
relevant to our analysis. This section also includes accounts of any 
peculiarities in our spectra that resulted in steps additional to those 
described above, notably NGC 1333 IRS 4A, Orion-KL, IRAS 05338-0624, 
NGC 2071 IR, and DR21(OH). Much has been published on these objects; 
readers seeking additional information should follow the citations below. 
	
\subsubsection{W3 IRS 4}
\label{w3irs4}
  
Embedded infrared source with associated compact HII Region 
W3(C) \citep{WW72}. Our spectra found two velocity components at 
-47.3 and -42.5 \kms{} in the \jthree{} transition, consistent with the \jone{} 
and \jtwo{} \f{} observations of \citet{Dickel96}, but failed to detect the 
\jfour{} transition. LVG modeling of several \f{} transitions from 211--365 GHz was employed by 
\citet{MW93} to find \textit{T}\tsb{K} = 75\pms{20}{15} K and 
log[\textit{n}(H\tsb{2})] = 5.40\pms{0.10}{0.20} \cm{}, while 
\citet{Helm94} found \textit{T}\tsb{K} = 55\pms{20}{10} K and 
\textit{n}(H\tsb{2}) = 10\tsp{6} \cm{} in a study using several molecules, including four \f{} transitions at around 364 GHz. 

\subsubsection{W3(OH)}
\label{w3oh}

Compact HII region with a shell structure \citep{DW81} surrounding an 
embedded infrared source with characteristics of a young O star 
\citep{Camp89} and extensive maser activity (e.g. \citealt{Mosc10}; 
Figure~\ref{fig:ch3oh} of this paper).  \citet{Mauer88} detected a warm molecular 
core and red-shifted emission feature using NH\tsb{3} with central 
velocity $\sim$ -47 \kms{}. Absorption features, red-shifted with 
respect to the main emission component, have also been observed 
in HCO\tsp{+} and OH \citep{Wink94,BM95}. The dominant emission 
and absorption components are detected by both our \jfour{} and 
\jthree{} spectra. The $J=3$ absorption profile exhibits an additional 
velocity component at 47.0\,km\,s$^{-1}$, which dominates the 2 and 6
cm \f{} observations 
of \citet{DG87}. The absence of this component from our $J=4$ 
spectrum is indicative of a foreground absorbing layer. \citet{MW93} 
used LVG modeling of several \f{} transitions from 211--365 GHz to find \textit{T}\tsb{K} 
= 110\pms{5}{15} K and log[\textit{n}(H\tsb{2})] = 5.95\pms{0.05}{0.01} 
\cm{} for the emitting material, while \citet{DG87} suggest that the \f{} 
absorption likely arises from material of density $\sim$ 5\e{4} \cm{} assuming
a temperature of 60 K based on the work of \citet{GSD83}. They note that
60 K likely represents an upper limit to the temperature of the \f{} absorption
region and also test 40 K to demonstrate the temperature dependency of their 
procedure. 

\subsubsection{L1448 IRS 3B}
\label{l1448}
		
3A and 3B are Class 0 sources separated by 6$\arcsec$ that share a 
common envelope in a potential protobinary system \citep{Bar98}, 
of which 3B dominates at mm wavelengths \citep{TP97}. 
H\tsp{13}CO\tsp{+} and N\tsb{2}H\tsp{+} observations by 
\citet{Volg06} indicate a systemic velocity between 3.4--5.8 \kms{}, 
consistent with our \jthree{} observation (\lsr{} = 4.8 \kms{}). 
The \jfour{} transition was not detected. \citet{Maret04} employed 
LVG modeling of several \f{} transitions from 141--364 GHz to find \textit{T}\tsb{K} 
= 90 K and \textit{n}(H\tsb{2}) = 10\tsp{5} \cm{}.	

\subsubsection{NGC 1333 IRAS 4A and B}
\label{n1333}
		
Premier examples of embedded, low mass star formation. Separated 
by $\sim$ 30$\arcsec$, 4A and B are the brightest of three sources 
in the IRAS 4 core \citep{Sand91} and each have independently 
been resolved into binary systems \citep{Loon98,Loon00}. P Cygni 
profiles, indicative of infall and characterized by a blueshifted emission 
component and redshifted absorption, have been detected toward 
both cores in interferometric observations of \f{} (3\tsb{12}-2\tsb{11}, 226 GHz) and CS \citep{DiF01}. 
This profile is eschewed in all but our \jthree{} observations of IRAS 
4A, which indicates an absorption component at 7.5 \kms{} with two 
emission components at $\sim$ 7.0 \kms{} being detected in both 
transitions. The broader, wing component of the emission can be 
attributed to outflow material. The absorption feature in the \jthree{} 
spectrum of IRAS 4A has the effect of bisecting the Gaussian profiles 
of both emission components, eliminating the high velocity component of the profile detected 
in the \jfour{} transition but without bringing the net $J=3$ profile below 
the baseline. Because of the blending in this spectrum, the total 
integrated emission over the FWZI of the entire line profile for 
IRAS 4A has been used for our analysis 
with the integrated intensity from the Gaussian fit to the absorption 
component added to the total $J=3$ intensity. 

The observations toward IRAS 4B were well fit by single Gaussians 
between 6.9 and 7.1 \kms{}, in agreement with the average systemic 
velocity of 7.0 \kms{} found for the IRAS 4 group by \citet{DiF01}. 
However, the \jthree{} emission was anomalously low, resulting in a 
transition ratio that could not be fit by the LVG model. Based on the 
aforementioned detection of \f{} absorption toward 4B within the velocity 
range over which our emission was observed, it is possible that the 
$J=3$ emission is being subtracted by an absorption component not 
visually reflected in our line profile. This possibility is discussed in detail 
in \S\ref{SourceSize}.

Using LVG modeling of several \f{} transitions from 141--364 GHz, \citet{Maret04} derived 
the following properties: 4A--\textit{T}\tsb{K} = 50 K, \textit{n}(H\tsb{2}) 
= 3x10\tsp{5} \cm{}; 4B--\textit{T}\tsb{K} = 80 K, \textit{n}(H\tsb{2}) 
= 3\e{5} \cm{}. \citet{Bla95} analyzed a combination of \f{} (several transitions from 218--365 GHz) and CS line 
ratios to find \textit{T}\tsb{K} = 20--40 K and \textit{n}(H\tsb{2}) = 
2\e{6} \cm{} for both core components, and \textit{T}\tsb{K} = 
70--100 K and \textit{n}(H\tsb{2}) = 5\e{6} \cm{} for the wing material 
near IRAS 4A.

\subsubsection{L1551 IRS 5}
\label{l1551}
		
Low mass star forming site suggested to be a binary system separated 
by $\sim$ 0.3$\arcsec$ \citep{Cohen84}. Single Gaussian emission 
profiles were detected in both transitions with a central velocity typical 
of the region at  $\sim$ 6.4 \kms{}. \citet{MS95} used \f{} (3\tsb{03}-2\tsb{02} and 3\tsb{22}-2\tsb{21}, 218 GHz) and CS 
observations to derive a kinetic temperature of $\sim$ 40 K and a spatial 
density of 6.9\e{6} \cm{}. \citet{Rob02} and \citet{RM07} examined transition ratios of several 
species, including the 2\tsb{11}-1\tsb{10} (150 GHz) and 5\tsb{14}-5\tsb{15} (72 GHz) transitions of \f{}, 
and quote a bolometric temperature of 97 K for 
the region. Because the \f{} transitions employed by \citet{MS95} are 
biased toward temperatures $\leq$ 50 K and the LVG prediction of our 
observed transition ratio supports a higher kinetic temperature, we have assumed 
100 K for our analysis while also testing temperatures of 40 and 150 K. 
Additional \f{} measurements have been conducted for L1551 for the 
\jone{} transition by \citet{Ara06}. The spatial density of IRS 5 has been 
estimated using several methods. \citet{But91} found an average volume 
density of 9\e{5} \cm{} within an angular radius of 13$\farcs$8 by 
modeling far-infrared emission, while \citet{Ful95} estimate 6\e{5} \cm{} 
using C\tsp{17}O emission. \citet{MS95} derived the significantly higher 
density of 11\e{6} \cm{} using transition ratios of the CS molecule. 

\begin{deluxetable*}{l c c c}
\tabletypesize{\scriptsize}
\tablecolumns{4}
\tablewidth{0pc}
\tablecaption{LVG Results: Velocity Components Detected in One Transition}
\tablehead{ 
\colhead{} & 
\colhead{\textit{T}\tsb{K}\tablenotemark{a}} & 
\colhead{log$\left[\frac{N(\small{\mathrm{ortho-\f{}}})}{\Delta\nu}\right]$\tablenotemark{b}} & 
\colhead{log[\textit{n}(H\tsb{2})]} 
\\ 
\colhead{Source} & 
\colhead{(K)} & 
\colhead{[cm\tsp{-2} (\kms{})\tsp{-1}]} & 
\colhead{(\cm{})} 
\\ \hline \\ 
\multicolumn{4}{c}{Emission Components} 
}
\startdata
L1448 IRS 3B & 90 & 13.17 & $\geq$5.32 \\
OMC-2 IRS 4$_{\mathrm{wing}}^{\mathrm{c}}$ & 85 & 14.27 & 4.93$\leq$\textit{n}$\leq$6.66 \\
Cep A East\tsb{wing} & 80 & 13.23 & $\geq$5.28  \\
NGC 7538 IRS 9 & 60 & 13.39 & $\geq$5.52 \\
\cutinhead{Absorption Components}
W3 IRS 4\tsb{peak} & 75 & 13.65 & $\leq$5.50 \\
W3 IRS 4\tsb{wing} & 75 & 13.17 & $\leq$5.36  \\
W3(OH) & 60 & 14.22 & $\leq$5.13  \\
NGC 1333 IRAS 4A & 50 & 13.71 & $\leq$5.52  \\
NGC 2024 FIR 5 & 25 & 13.99 & $\leq$5.58 \\
NGC 2024 FIR 6 & 25 & 13.21 & $\leq$5.84  \\
NGC 2071 IR & 80 & 13.48 & $\leq$5.38 \\
NGC 7538 IRS 1 & 25 & 14.43 & $\leq$5.17  \\
\enddata
\tablenotetext{a}{Culled from the literature. See \S\ref{comparisons} for references.}
\tablenotetext{b}{\textrm{L}TE approximation.}
\tablenotetext{c}{For this source only, the 3$\sigma$ detection limit could be used to determine upper and lower limits.}
\label{tab:LVGNondetections}
\end{deluxetable*}

\subsubsection{Orion-KL}
\label{orikl}
  
 The best-studied region of massive star formation to date (see review 
 by \citealt{GS89}) and extremely bright in a plethora of chemical species 
 (brightness temperatures from Orion-KL are $>$ 5$\times$ that of any other source 
 in our study). This is the only source in our sample for which previous 
 observations of the \jthree{} transition of \f{} have been made \citep{Wil80,MB80,Bas85}. 
 Outflows, shocks, and turbulence arising from newly formed stars in the region 
 have led to a complex velocity structure comprised of several distinct components 
 \citep{Blake87}, at least three of which are captured in our observations: 
 the hot core (\lsr{} $\sim$ 6 \kms{}, $\Delta\nu >$ 10  \kms{}, T\tsb{K} $\sim$ 
 300 K), compact ridge (\lsr{} $\sim$ 8 \kms{}, $\Delta\nu$ $\sim$ 4  \kms{}, 
 T\tsb{K} $\sim$ 135 K), and extended ridge (\lsr{} $\sim$ 9--10 \kms{}, 
 $\Delta\nu$ $\sim$ 4  \kms{}, T\tsb{K} $\sim$ 135 K), each with spatial density 
 \textit{n}(H\tsb{2}) $\gtrsim$ 10\tsp{5} \cm{} \citep{Man93,MW93}.
 
Our $J=4$ spectrum exhibits an anomalously intense velocity component at 
\lsr{} $\sim$ 10.3 \kms{}, which also appears as a much smaller contribution 
in the $J=3$ transition. A feature at this velocity exists 10$\arcsec$ north of our observed position \citep{Man90}, 
leading us to suspect a pointing error. Pointing was checked prior to observing, and the 
Orion-KL scan was followed by observations of OMC-2 IRS 4 and NGC 2024. 
If a pointing error is to be blamed, it should manifest itself in the subsequent 
observations of OMC-2 IRS 4 and NGC 2024, but we could find no evidence 
for this. The possibilities of an unidentified line or rest frequency error were 
also ruled out. This spectrum is the average of only two (mutually consistent) 
scans conducted during a single run. We suggest that this 
emission arises from the extended ridge, which is typically observed around 
\lsr{} $\sim$ 9 \kms{}, but occasionally as high as 10 \kms{}. This 
interpretation, and the physical parameters derived from it, should be 
applied cautiously for the reasons explained above. Because of the disparity 
between the relative intensity of the 10.3 \kms{} feature in each transition, 
only the transition ratio for the hot core component could be fit by the LVG model, and 
the extraordinarily high intensities precluded the determination of useful 
density limits for the other components through the procedure described in
\S\ref{LTE}. 

\subsubsection{Orion-S}
\label{oris}
		
Broad SiO emission coupled with relatively weak SO\tsb{2} and CH\tsb{3}OH 
emission indicates the presence of an energetic outflow in its earliest phase, 
suggesting that Orion-S is one the youngest stellar objects in the region 
\citep{Mc93}. Emission from \lsr{} $\sim$ 5.5--7.5 \kms{} has been 
observed in numerous dense gas tracers, including \f{} by \citet{Bas85} 
(2\tsb{11}-2\tsb{12}, 14 GHz) and \citet{Man90} (several transitions from 218--291 GHz). 
Additionally, \f{} absorption has been detected for the \jone{} transition by 
\citet{John83} over the velocity range 3.7--9.8 \kms{}, which \citet{Man93} 
also detect and attribute to a lower density region just north of the emission features. 
Using NH\tsb{3} data and the \jtwo{} observations of \f{} that would later 
be presented by \citet{Bas85}, \citet{Bat83} found a kinetic temperature 
of 100 K for the 6.5 \kms{} component and a density of 4\e{6} \cm{}, 
significantly lower than our result. However, they also detect a separate 
velocity component within our detected velocity range at 7.4 \kms{} for 
which they derive a density of 6\e{7} \cm{}, significantly higher than 
our result. 

We observe both the $J=4$ and $J=3$ transitions in emission with peaks at 
$\sim$ 6.4 \kms{}. Both spectra also exhibit broad wing components 
likely arising from the outflow material. The transition ratio for the wing emission 
was anomalously low and thus the density could not be fully constrained 
by the LVG model. Given that \f{} absorption has been observed over 
this velocity range, it is possible that the \jthree{} emission is being 
partially absorbed without an absorption component being visually 
reflected in the line profile. This possibility is discussed in detail in \S\ref{SourceSize}. 

\begin{deluxetable}{l c c c}
\tabletypesize{\scriptsize}
\tablecolumns{4}
\tablewidth{0pc}
\tablecaption{LVG Results: Single Transition Observations}
\tablehead{ 
\colhead{} & 
\colhead{} & 
\multicolumn{2}{c}{log$\left[\frac{N(\small{\mathrm{ortho-\f{}}})}{\Delta\nu}\right]$\tablenotemark{b}} 
\\ 
\colhead{} & 
\colhead{\textit{T}\tsb{K}\tablenotemark{a}} & 
\multicolumn{2}{c}{[cm\tsp{-2} (\kms{})\tsp{-1}]} 
\\ 
\colhead{Source} & 
\colhead{(K)} & 
\colhead{LVG} & 
\colhead{LTE}
}
\startdata
Orion-N & 35 & 14.21\pms{0.04}{0.02} & 13.95$\pm0.02$ \\
NGC 2024 & 95 & 13.81\pms{0.09}{0.11} & 13.95\pms{0.10}{0.13}\\
G34.26+0.15 & 160 & 13.80\pms{0.14}{0.21} & 14.20\pms{0.12}{0.19} \\
S68N & 35 & 13.88\pms{0.09}{0.12} & 13.58\pms{0.13}{0.18} \\
W51M & 150 & 14.03\pms{0.13}{0.17} & 14.40\pms{0.12}{0.17} \\
\enddata
\tablenotetext{a}{Culled from the literature. See \S\ref{comparisons} for references.}
\tablenotetext{b}{\textrm{L}VG column density derived assuming \textit{n}(H\tsb{2})=10\tsp{6.15} \cm{}, the average density from Table~\ref{tab:LVGDetections}. LTE results from Table~\ref{tab:h2coMeas} provided for comparison.}
\label{tab:LVGSingleLine}
\end{deluxetable}

\subsubsection{OMC-2 IRS 4}
\label{omc2irs4}
		
Infrared source composed of two objects designated 4N and 4S, 
which are separated by $\sim$ 4$\arcsec$ \citep{Pen86} and 
associated with the extended (29$\arcsec$x13$\arcsec$) source 
FIR 3 \citep{Mez90}. Our observations exhibit Gaussian profiles 
consistent with the literature and peaked at $\sim$ 11.5 \kms{}. 
The $J=3$ profile includes an additional wing component that can be attributed 
to an outflow in the region. Previous LVG modeling of several \f{} transitions from 211--365 GHz 
was conducted by \citet{MW93}, who found \textit{T}\tsb{K} = 
85 K and log[\textit{n}(H\tsb{2})] = 5.65 \cm{}, nearly identical 
to our result. \citet{Mez90} estimated the density of OMC-2 over 
a 50$\arcsec$x50$\arcsec$ region to be 1.9\e{6} \cm{} using 
inferred mass and source size arguments. 

\subsubsection{IRAS 05338-0624}
\label{i05338}
		
Young stellar object associated with L1641-N, a cluster of infrared 
sources \citep{Strom89,Chen93} found to be mainly low-mass, 
pre-main sequence stars \citep{HD93}. \citet{Chen96} used an 
analysis of CS observations to estimate the density of the molecular 
core to be $\sim$ 10\tsp{6} \cm{}, while \citet{Mc94} also used 
CS to find a density of 2\e{5} \cm{} assuming a temperature of 
42 K derived from dust continuum observations \citep{Wal90}. 
\citet{SW07} uncovered higher temperatures ($\sim$ 150 K) 
within a 1$\farcs$4 region using CH\tsb{3}CN observations. 
Since the dust continuum represents, at best, a lower limit to 
the kinetic temperature of the gas and the high temperatures found by 
\citet{SW07} are constrained to an area much smaller than 
our beam size, we have used an average of 95 K for our analysis 
and tested both extremes.

Both of our \f{} spectra exhibit a two-component structure 
with a central peak at  \lsr{} $\sim$ 7.1 \kms{} and a shoulder 
profile centered around \lsr{} $\sim$ 9.6 \kms{}. The shoulder 
profile is attributed to the red lobe of the outflow material and is 
consistent with the spectra of several species observed by 
\citet{Mc94}. Because the shoulder profile is very weakly detected 
in the $J=4$ profile, our confidence in the Gaussian fitting routine's 
ability to reliably separate the two components was low, so the total 
integrated emission over the FWZI of the entire line profile was used in our analysis. 

\subsubsection{NGC 2024 FIR 5 and 6}
\label{n2024}
  
Star forming region containing a string of dense cores, FIR 1-7, 
embedded in an extended molecular ridge \citep{Mez92}. Both 
transitions were detected toward FIR 5 and 6, and an additional \jfour{} 
measurement was made for the ridge material (approximately midway 
between FIR 4 and 5). The following properties have been derived from 
LVG modeling of several \f{} transitions (211--365, 632 GHz): FIR 5--\textit{T}\tsb{K} = 
160\pms{90}{60} K, \textit{n}(H\tsb{2}) = 1$\pm$0.5\e{6} \cm{}; FIR 
6--\textit{T}\tsb{K} = 40\pms{120}{10} K, \textit{n}(H\tsb{2}) = 
2$\pm$0.5\e{6} \cm{}; Ridge--\textit{T}\tsb{K} = 95\pms{30}{20} 
K \citep{MW93,Man99,WM08}.
 
The absorption feature exhibited by both of our \jthree{} observations 
is attributed to a cool (20--30 K) foreground layer, which has been 
observed in the \jone{} and \jtwo{} transitions of \f{} 
and found to have \textit{n}(H\tsb{2}) = 10\tsp{4.9} \cm{} \citep{Hen82,Cru86}. The physical 
parameters of NGC 2024 have been explored using a variety of other tracers 
as well, recently CO by \citet{Emp09}, who found the bulk of the material to 
be characterized by \textit{T}\tsb{K} $\sim$ 75 K and \textit{n}(H\tsb{2}) 
$\sim$ 9\e{5} \cm{}.

\subsubsection{NGC 2071 IR}
\label{n2071ir}
		
Cluster of infrared sources spanning $\sim$ 30$\arcsec$ \citep{Per81}; 
our observations are centered on IRS 1, but IRS 2 and 3 are also being 
sampled. Both spectra indicate a central velocity consistent with the 
literature at $\sim$ 9.7 \kms{}. An additional shoulder component arising 
from dense gas in the outflow was also detected in both transitions at $\sim$ 
8.3 \kms{}, and an absorption component was observed in the $J=3$ profile 
at $\sim$ 16.2 \kms{}. The 16 \kms{} feature has been previously observed 
in emission for the CS \textit{J} = 1--0 line by \citet{Tak84} and \citet{Kit02}. 

Previous \f{} 
measurements of several transitions from 211--365 GHz by \citet{MW93} 
estimate \textit{T}\tsb{K} = 80 K and \textit{n}(H\tsb{2}) = 10\tsp{6} \cm{}. 
\citet{Tau88} used the 3\tsb{13}-2\tsb{12} (211 GHz) and 3\tsb{12}-2\tsb{11} (225 GHz) transitions of \f{} 
to estimate the density in the region and 
found \textit{n}(H\tsb{2}) $\sim$ 3\e{5} \cm{}. A microturbulent model of 
CS and C\tsp{34}S emission from \citet{Zh91} yielded a best fit to the 
density at 3\e{6} \cm{} with the density of the outflow emission being 
$>$ 10\tsp{5} \cm{}. Our LVG analysis yielded a significantly higher 
density for the outflow component than for the central emission peak. 
It is possible that our two-component Gaussian fit to the spectra is improperly 
separating the components. An analysis of the total integrated emission 
over the FWZI of the entire line profile yields \textit{n}(H\tsb{2}) = 10\tsp{6.00} \cm{}.

\subsubsection{S255N}
\label{s255n}
		
Massive star-forming region also known as S255 FIR 1 or G192.60-MM1 
that lies at one end of an extended molecular ridge opposite S255IR 
\citep{Hey89}. Three compact cores, SMA1--3, were resolved in the 
1.3 mm continuum maps of \citet{Cyg07}, who also combined an 
analysis of \f{} transition ratios (3\tsb{03}-2\tsb{02}, 3\tsb{22}-2\tsb{21}, and 3\tsb{21}-2\tsb{20}; 218--219 GHz) 
and a spectral energy distribution (SED) 
model to estimate a temperature range of 40--100 K and densities 
between 3--16\e{6} \cm{}. The \f{} observations presented in \citet{Cyg07} 
are centered on either side of SMA1 (NE and SW) and indicate velocity 
components at \lsr{} $\sim$ 6.9 \kms{} and 12.1 \kms{}, while our spectra, 
centered between the two, show a single component at $\sim$ 8.9 \kms{}. 

\subsubsection{G34.26+0.15}
\label{g34}

Ultra compact (UC) HII region with an associated hot core that has 
become a prototypical example of cometary morphology \citep{WC89,vB90}. 
The hot molecular gas (80--175 K) is suggested to be the outer layer 
of a massive, cool core that is being externally heated by the UC HII 
region, from which it is offset by $\sim$ 2$\arcsec$ \citep{Heat89,WM99}. 
Single dish CH\tsb{3}CN observations conducted by \citet{Chu92} 
suggest gas temperatures of 166 K and densities $>$ 10\tsp{5} \cm{}, 
while \citet{Mook07} also found a temperature of 160 K using the 
brightness temperatures of several optically thick lines.

\subsubsection{S68N}
\label{s68n}
	
Deeply embedded source within the Serpens molecular cloud containing
 a protostar with an associated outflow \citep{Wolf98}, suggested by 
 \citet{Mc00} to be an example of a ``cool core,'' with properties intermediate 
 to cold/warm condensations. \citet{Mc00} used relative intensities of four \f{} 
 transitions [1\tsb{01}-0\tsb{00} (73 GHz), 2\tsb{02}-1\tsb{01} (146 GHz), 3\tsb{03}-2\tsb{02} (218 GHz), and 3\tsb{22}-2\tsb{18} (218 GHz)] 
 to derive a kinetic temperature range of 35--70 K but report 
 flat-topped line profiles suggesting an overestimation due to optical depth 
 effects and adopt an estimate of 35 K for their measurements. \cite{Hurt96} 
 also examined four \f{} transitions [3\tsb{03}-2\tsb{02} (218 GHz), 3\tsb{22}-2\tsb{18} (218 GHz), 5\tsb{05}-4\tsb{04} (363 GHz), 5\tsb{23}-4\tsb{22} (365 GHz)] to determine a higher temperature of 
 around 75 K. Both studies also report spatial density estimates with \citet{Mc00} 
 finding \textit{n}{(H\tsb{2}) = 0.4--1.2\e{6} \cm{} using H\tsp{13}CO\tsp{+}, 
 SiO, and DCN, while \citet{Hurt96} found 2.5\e{6} \cm{} using \f{}.

\subsubsection{W51M}
\label{w51m}

Dominant region of the massive star-forming site W51, which \citet{Mar72} 
revealed to consist of eight distinct components at centimeter wavelengths. 
Our observations are centered on W51e, itself divided into four UC HII 
regions e1--4 \citep{Gau93}, whose association with dense molecular cores 
and maser activity is known as W51-Main \citep{ZH97}. \citet{Rem04} used 
CH\tsb{3}CN to find temperatures of 123 and 153 toward e1 and e2, respectively, and 
a density of 5\e{5} \cm{} for both sources. \citet{ZH97} derive higher densities 
of 2--3\e{6} \cm{} for the region using NH\tsb{3}. 

\subsubsection{DR21(OH)}
\label{dr21oh}

Star-forming region in an early phase of its evolution which has yet to see 
substantial ionization of the molecular material surrounding newly-formed, 
massive B stars. \citet{Man92} detected four principal condensations labelled 
M, N, W, and S. Our observations are centered on DR21(OH)-M, which is 
subdivided into two components separated by $\sim$ 8$\arcsec$, MM1 and 
MM2. Despite their proximity, MM1 and MM2 differ substantially. MM1 is hot 
($\sim$ 160 K), dense ($\sim$ 10\tsp{6} \cm{}), and moderately luminous 
($\sim$ 10\tsp{4}\lsolar{}), while MM2 is cooler ($\sim$ 30 K), denser 
($\sim$ 10\tsp{7} \cm{}), and less luminous ($\sim$ 10\tsp{3}\lsolar{}) 
\citep{Man91,Man92,MW93}. Our $J=3$ spectrum indicates distinct peaks 
at -4.5 and -1.3 \kms{}, attributed to MM1 and MM2, respectively. The $J=4$ 
spectrum is severely blended and demands a surprisingly high contribution 
from the -1.3 \kms{} (MM2) component. Since this region has not been 
mapped at these frequencies and the spectra were so highly blended, 
due caution should be exercised when assessing our results.

\subsubsection{Cep A East}
\label{cepa}

Massive star forming region known to consist of 16 compact 
($\sim$ 1$\arcsec$) components clustered within a 25$\arcsec$ 
radius and aligned in an inverted Y-shaped structure \citep{Gar96}. 
Our observations are centered on the continuum source HW3, 
$\sim$ 3$\arcsec$ south of the dominant source HW2. HW2 
is well within our beam and is associated with a promising candidate 
for the detection of a massive disk \citep{Tor96,Pat05}, though 
other studies suggest that this elongated molecular structure is explained 
by the superposition of at least three hot cores \citep{Com07,Bro07}. 
Choosing an appropriate kinetic temperature estimate was difficult for 
this region because of the inclusion of several compact sources within 
our beam. We have assumed a range of 40--100 K based on the estimates 
of \citet{Bro07} and \citet{Cod05} for the ambient cloud velocity of $\sim$ 10.5 \kms{}. 
\citet{Cod05} also find densities between 2--6\e{6} \cm{} in the region 
using SO and SiO. 

\subsubsection{NGC 7538 IRS 1}
\label{n7538i1}

The brightest of three compact sources discovered by \citet{Mar73}, 
which contains a zero-age main sequence star of spectral type earlier 
than O7.5 \citep{Lug04} embedded in a molecular cloud with an inner 
shell structure surrounding the star \citep{Pra97}. A systemic velocity 
of around -57 \kms{} was observed in several transitions of \f{} between 365--470 GHz by \citet{Tak00} with an 
absorption feature near -60 \kms{} having been detected in NH\tsb{3} 
\citep{Wil83}, both of which are found in our spectra. The region is 
also notable for the rare occurrence of 6 cm (\jone{}) \f{} masers 
\citep{Rot81,Hof03}. The absorption component we observe was 
also detected by \citet{Hof03} in the 2cm (\jtwo{}) transition of \f{}.

A temperature of $\sim$ 220 K was found using NH\tsb{3} by 
\citet{Mauer88}, while a cooler temperature of 176 K was 
found by \citet{Mit90} using \tsp{13}CO and a warmer one 
of 245 K by \citet{Qiu11} using CH\tsb{3}CN. The Mitchell study also 
indicates a cold gas component of 25 K from which the absorption 
component in our $J=3$ spectrum likely arises. From \tsp{13}CO observations, 
\citet{Qiu11} estimate the density to be $\sim$ 10\tsp{7} \cm{} 
for a region about half the size of our beam, while \citet{Mit90} 
indicate that the density is $>$ 10\tsp{6} \cm{}. \citet{Hof03}
note that to excite the \jone{} \f{} masers in the region, 
lower densities in the range of 6--16\e{4} \cm{} are required. 
The masers arise from a region directly in front IRS 1 not coincident
with the hot core region examined by the previously mentioned 
authors. 

Since our observations are of higher excitation transitions than the
\f{} masers in the Hoffman study, it is likely that we are sampling material
from both the hot core and maser regions, and thus the intermediate 
density we find is justified. However, this notion also calls into question
the validity of our kinetic temperature assumption, so it should be noted
that temperatures $<120$ K would result in densities $>$ 10\tsp{6} 
in the LVG approximation. That said, due to the excitation requirements of the 
\jthree{} and \jfour{} transitions compared to that of the $J=1$ \textit{K}-doublet 
observed by \citet{Hof03}, it is more likely that our measurements 
are biased toward the higher temperatures of the hot core. Fortunately, the 
\f{} LVG models are relatively independent of \textit{T}\tsb{K} for temperatures 
$\gtrsim100$ K (see \S\ref{temp}).

\subsubsection{NGC 7538 IRS 9}
\label{n7538i9}

Deeply embedded cold IR source $\sim$ 50$\arcsec$ south of IRS 1 
that is associated with a large reflection nebula \citep{Wer79}. 
The radial velocity of the IRS 9 cloud core from H\tsp{13}CN, 
CS, and \f{} data is about -57 \kms{} \citep{San05,Tak00}, consistent 
with our observations. Single-dish JCMT \f{} 
(3\tsb{22}-2\tsb{21} and 3\tsb{21}-2\tsb{20}, 218 GHz) spectra suggest a gas 
temperature of $\geq$ 60 K \citep{San05}, while HCO\tsp{+}, 
H\tsp{13}CO\tsp{+}, and \tsp{13}CO observations suggest a lower 
temperature of $\sim$ 30 K \citep{HM95,Mit90}. \citet{Mit90} also 
detect the presence of warm gas (180 K) and suggest that the 
density of this material is $>$ 10\tsp{6} \cm{}. 

\section{Limitations of \f{} \textit{J} = 3/\textit{J} = 4 \textit{K}-Doublet Densitometry}
\label{discussion}

Our study is concerned primarily with the application of a previously 
unused densitometry technique and, as such, well-studied objects were
chosen to assess the method's performance. This discussion is focused
on limitations to the technique that
require elaboration. These effects may also help explain the problem of 
the anomalously low transition ratios [$\int{}T_{mb}(\jthree)d\nu$/$\int{}T_{mb}(\jfour)d\nu$]
 that precluded LVG modeling in a few
sources (see Table~\ref{tab:LVGDetections}).

\begin{deluxetable}{l c c}
\tabletypesize{\scriptsize}
\tablecolumns{3}
\tablewidth{0pc}
\tablecaption{Temperature Dependence of LVG Density Derivation}
\tablehead{ 
\colhead{$T_K$} & 
\colhead{log[\textit{n}(H\tsb{2})]\tablenotemark{a}} & 
\colhead{\% Difference} 
\\ 
\colhead{(K)} & 
\colhead{(\cm{})} & 
\colhead{Over $T_K$ Range} 
}
\startdata
30--50 & 7.72--6.71 & 923 \\
50--100 & 6.71--6.12 & 290 \\
100--150 & 6.12--5.91 & 59.2 \\
150--200 & 5.91--5.81 & 26.2 \\
200--250 & 5.81--5.75 & 15.0 \\
250--300 & 5.75--5.69 & 15.0 \\
\enddata
\tablenotetext{a}{Calculated for test case NGC 7538 IRS 1.}
\label{tab:Tdepend}
\end{deluxetable}

\subsection{LVG Model Dependence on Kinetic Temperature}
\label{temp}

The LVG model for the \jthree{} and \jfour{} transitions is 
somewhat dependent on kinetic temperature (\textit{T}\tsb{K}). As previously 
mentioned in \S\ref{LVG}, this meant that a careful selection of appropriate
kinetic temperatures from previous studies was necessary. Wherever possible, 
temperatures derived from \f{} measurements were used to ensure coupling to 
the gas traced by our observations. Estimates taken from analyses of other 
dense gas tracers such as NH\tsb{3} or CH\tsb{3}CN are otherwise 
preferable. That said, any time a temperature derived from 
measurements of disparate molecules (or even \f{} measurements of significantly 
different excitation requirements) is adopted, the question of whether or not this estimate 
can be associated with the gas sampled by our beam must be raised. In at least 
one case (NGC 7538 IRS 1, \S\ref{n7538i1}) there appears to be a distinct possibility that the 
temperature we adopted may not be wholly appropriate for the material traced by our 
observations. 

In Table~\ref{tab:LVGDetections}, we elected to include the spatial densities derived 
for a range of kinetic temperatures based on the error estimates in our assumed values.
This is because the temperature dependence of our \f{} LVG models is not linear and itself 
depends on the range of physical parameters being studied. Table~\ref{tab:Tdepend} 
provides a summation of this dependence by dividing the range 
of kinetic temperatures found in our sample into 6 groups. The test case of NGC 7538 IRS 1 
was chosen for its moderate transition ratio (R$_i$) of 1.20. From the table it is clear that the dependence on \textit{T}\tsb{K} 
decreases with an increase in the assumed kinetic temperature.
In the most volatile range (30--50 K) the measured density can fluctuate by upwards of an 
order of magnitude between the extremes. Fortunately, very few of our sources fall in this range. At high 
temperatures ($\gtrsim200$ K) the dependence on \textit{T}\tsb{K} is fairly constant. In this regime, uncertainties in the 
derived density due to errors in the adopted kinetic temperature are generally below the uncertainty 
imposed by the collisional excitation rates between \f{} and H\tsb{2} (20\%; see \S\ref{LVG}). 

Figure~\ref{fig:LVGTemp} 
provides a graphical representation of the \f{} LVG model's dependence on \textit{T}\tsb{K} with plots identical 
to that of Figure~\ref{fig:LVG} for a source of high and low kinetic temperature. Note how at high temperatures (upper panel), 
the transition ratio and brightness temperature contours approach orthogonality. In effect, this means that as the kinetic 
temperature varies and the locations of the curves change with respect to each other, the intersection point between 
them is moved by smaller amounts at higher temperatures. For \textit{T}\tsb{K}  $\gtrsim200$ K, the brightness 
temperature (solid-line) contours become roughly horizontal and the solution point is therefore only affected 
by the lateral motion of the transition ratio contours. This is why a variation in \textit{T}\tsb{K} from 200--250 K 
produces the same variation in spatial density as 250--300 K (see Table~\ref{tab:Tdepend}).

It should also be noted that, depending on the observed $J=3/J=4$ integrated intensity (line) ratio, the kinetic temperature 
assumption can be the deciding factor in whether or not the LVG model can successfully 
match an observation. At high temperatures, the higher excitation $J=4$ transition can start 
to enjoy a greater relative population than the $J=3$, and R$_i<1$ can reasonably 
be expected. In Table~\ref{tab:LVGDetections} there are 3 sources, L1551 IRS 5, Orion-S, and 
DR 21(OH)-MM1, with transition ratios on order $\sim0.9$. This was not a problem for the assumed 
temperature, but at the lower temperature limit, the density is indicated to be $>10^8$ (column 5). 
This is to say that the LVG prediction for the density is unbound at such low temperatures. In terms of 
Figure~\ref{fig:LVGTemp}, for kinetic temperatures below a given value, the transition ratio (color) and brightness 
temperature (solid-line) contours no longer intersect (as in Cep A East from Figure~\ref{fig:LVG}).

Given this discussion and the uncertainties typical in estimates of kinetic temperature, we 
suggest that \f{} \jthree{}/\jfour{} densitometry is best suited to objects 
with \textit{T}\tsb{K} $\gtrsim$ 100 K. We have demonstrated that this technique is also 
viable at lower kinetic temperatures but with the caveat that its results become increasingly 
sensitive to uncertainties in the assumed value of \textit{T}\tsb{K}. The reason for this 
effect can likely be attributed to the relative excitation requirements of the $J=3$ and 4 transitions, 
which are discussed in the following section.

 \begin{figure}
 \figurenum{5}
 \epsscale{.9}
 \plotone{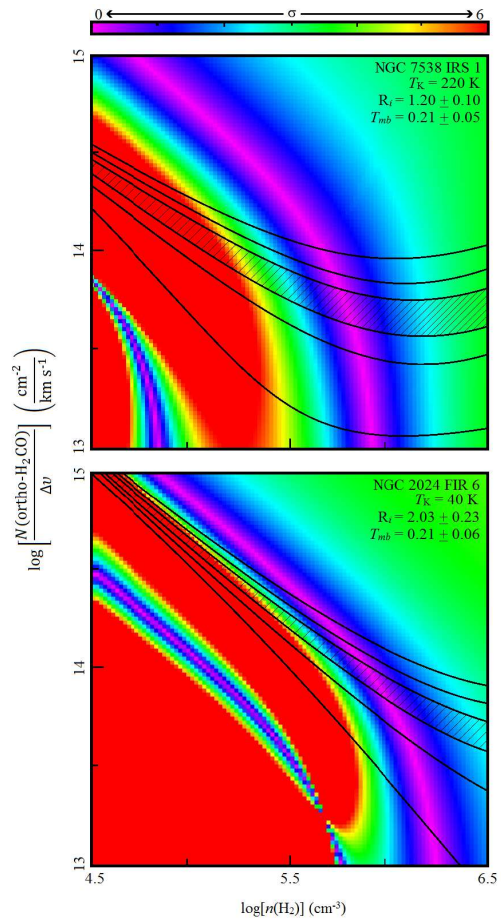}
 \caption{Comparison of model results for sources of high (upper panel) and low (lower panel) 
 kinetic temperature. Contours and parameters are as in Fig.~\ref{fig:LVG}. Note how at high \textit{T}\tsb{K}, 
 contours approach orthogonality, resulting in less variation in the solution point due to changes 
 in \textit{T}\tsb{K} (alternatively, less dependence on \textit{T}\tsb{K}).}
 \label{fig:LVGTemp}
 \end{figure}

\subsection{\f{} \jthree{} and \jfour{} Relative Excitation Requirements}
\label{exeffects}

In addition to the kinetic temperature of an object, it is important to consider the
relative excitation requirements for each transition. Being of lower excitation, 
the \jthree{} transition can be excited by temperatures and 
densities lower than that required by the \jfour{} transition. This means that the 
$J=3$ transition is predictably observed to be more intense 
than the $J=4$ in the general case. What complicates the matter is that the two 
transitions are affected differently by the collisional pumping mechanism described 
in \S\ref{h2coProbe}. In effect, this means that the \jfour{} transition falls into absorption at a 
slightly higher spatial density than the \jthree{}. Therefore a small 
band of parameter space, briefly described in \S\ref{LTE}, exists for which the $J=4$ transition
is in absorption while the $J=3$ is in emission. 

It may be counterintuitive, then, that only one absorption component was detected in the \jfour{} 
transition while 9 were observed in the \jthree{}. This is explained by two effects. The first is that 
the system temperatures for the $J=4$ (Q-band) observations were substantially higher than the 
$J=3$ (Ka-band), resulting in much noisier \jfour{} spectra. Looking at the spectra in Figure~\ref{fig:DualDetections}, 
the majority of absorption components detected in the \jthree{} transition are below the noise of the \jfour{} spectra. 
The second, and perhaps more significant, effect is that the $J=3$ beam is substantially larger than that of the $J=4$, 
meaning that the material being sampled is a bit different for each transition. This topic is addressed in \S\ref{SourceSize}. 

Figure~\ref{fig:tex} displays the excitation temperatures of the $J=3$ and 4 transitions as a function of spatial density for an 
ortho-\f{} column density roughly equivalent to the average of that observed in our sample. From this, we can see the 
effect of the collisional pumping mechanism described in \S\ref{h2coProbe}, which cools the excitation temperatures of the 
$J\leq5$ \textit{K}-doublet transitions to lower than 2.73 K for a given set of physical conditions, allowing them to absorb 
radiation from the cosmic microwave background. Curves for 
kinetic temperatures of 50, 150, and 200 K are shown. Note how at high kinetic temperatures, the excitation temperatures for both 
transitions match more closely and only barely fall below the 2.73 K level. This means that at high 
kinetic temperatures, the possibility of observing absorption is small and the difference between the excitation requirements of each 
transition is of less consequence. This is likely the reason why the \jthree{}/\jfour{} densitometry technique is best suited for 
temperatures $\gtrsim$ 100 K.  

\begin{figure}
 \figurenum{6}
 \includegraphics[angle=-90, width=0.45\textwidth]{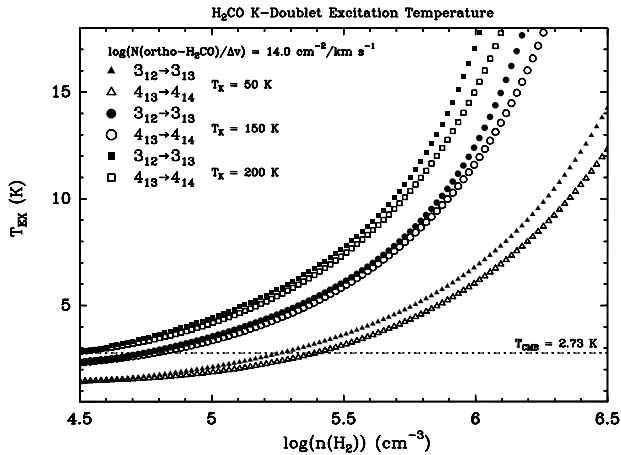}
 \caption{\f{} $J=3$ and 4 \textit{K}-doublet excitation temperatures as a function of molecular hydrogen density for an 
 ortho-\f{} column density per velocity gradient roughly equivalent to the average value observed in our sample. Note that the 
 possibility for absorption is much smaller at high kinetic temperatures than it is for low. Values predicted via LVG modeling.}
 \label{fig:tex}
 \end{figure}

\subsection{The Effects of Spatial Extent}
\label{SourceSize}

Because the beam sizes differ by
$(\theta_b(29\,GHz)-\theta_b(48\,GHz))/\theta_b(48\,GHz) = 63\%$, it is  
important to consider the distribution of dense gas in each object as the larger 
beam size observations ($J=3$) may be sampling extended structure
that biases the resulting line ($J=3/J=4$ integrated intensity) ratio. In similar studies using the \jone{} and 
\jtwo{} \textit{K}-doublets, the beam sizes of both observations 
(153$\arcsec$ and 51$\arcsec$, respectively, for the GBT) can generally be
considered much larger than the dense gas distribution and it is
safe to assume that all relevant emission is sampled. One
advantage of the \jthree{} and \jfour{} transitions is that their smaller
beam sizes (26$\arcsec$ and 16$\arcsec$, respectively, for the GBT) allow for higher 
resolution and heightened sensitivity to the spatially compact regions 
in which stars are known to form. However, these beam sizes also
approach the spatial extent of molecular cores and it is therefore prudent 
to consider potential differences in the material being sampled by each observation.

Unfortunately, no mapping measurements of the \f{} $J=3$ and 4 
\textit{K}-doublet transitions exist. We experimented with assuming spatial extents derived from 
other dense gas tracers in the literature but 
ultimately decided that inconsistencies in the frequency, chosen tracer, excitation 
conditions, and spatial resolution precluded truly appropriate spatial extent estimates for our purpose.
Instead, source sizes equal to the smaller beam size (16$\arcsec$) were
assumed and the $J=3$ observations were scaled for beam dilution (see
\S\ref{observations}). We find this agreeable because 16$\arcsec$ is a
reasonable average value for estimates of the dense gas distribution in
most of our sources, but this is surely not appropriate for each application. 

It is also possible for an object's dense gas distribution to be significantly 
smaller than 16$\arcsec$, in which case the $J=4$ observation is also 
subject to beam dilution. To examine this effect, consider the case of 
L1551 IRS 5, for which high-resolution imaging of CS suggests
the dense gas to be distributed over $\sim7\arcsec$ \citep{Tak04}. Applying the 
beam dilution correction described in \S\ref{observations} to this case has the 
effect of boosting the intensity of both transitions, but comparatively more for 
the $J=3$ observation, resulting in an increased transition ratio [$\int{}T_{mb}(\jthree)d\nu$/$\int{}T_{mb}(\jfour)d\nu$]. The subsequent LVG 
approximation to the density (10\tsp{5.95} \cm{}) is nearly an order of magnitude 
lower than that of the 16$\arcsec$ assumption (10\tsp{6.87} \cm{}). This decrease
is mainly due to an increased optical depth in the modelled transitions. 

In several sources, the transition ratio was anomalously low (see 
Table~\ref{tab:LVGDetections}), meaning that the $J=3$ transition was observed to be 
unexpectedly less intense than the $J=4$. This is possible at temperatures and 
densities high enough to populate the higher excitation transition significantly more than the lower, 
but these conditions are not met by the sources in question. One possible explanation 
for this effect is that the $J=3$ transition is being subjected to absorption that is not reflected 
in the line profile other than to lower its intensity. The LVG model assumes a uniform spatial density, which proves to 
be a reasonable approximation but ignores a density gradient that likely exists in molecular 
cores, with the outer regions being comparatively less dense than the inner. 
It is possible that the larger $J=3$ beam is sampling the comparatively cool, diffuse material 
enveloping a molecular core which has resulted in some self absorption. Indeed, in a few sources (e.g. NGC 1333 IRAS 4B and Orion-S, \S\ref{n1333} 
and \S\ref{oris}, respectively), \f{} absorption has been detected in the $J=2$ or 1 
\textit{K}-doublet transitions over the velocity range in question. Mapping studies are required to disentangle 
the varying density structure in many of the sources in our sample. 
	
\section{Conclusion}
\label{conclusion}

Using observations of the \jthree{} and \jfour{} 
transitions of \f{}, we have successfully constrained the spatial densities 
of a sample of galactic star-forming regions. Both transitions were observed 
toward 18 sources with relative ease, requiring an average of 17 min of 
integration time and resulting in only 3 nondetections in the $J=4$ 
transition. Accurate measurements of the spatial density 
[\textit{n}(H\tsb{2})] were made for 13 objects and useful limits 
were placed on the remainder using a combination of Large 
Velocity Gradient (LVG) and Local Thermodynamic Equilibrium (LTE) 
analyses. Molecular hydrogen densities in the range of 10\tsp{5.5}--10\tsp{6.5} \cm{} 
and ortho-formaldehyde column 
densities per unit line width between 10\tsp{13.5} and 10\tsp{14.5} cm\tsp{-2} (\kms{})\tsp{-1} are found for most sources, in general
 agreement with previous measurements. 

Detailed analyses of the advantages and limitations to this densitometry technique have also been provided. 
\f{} \jthree{}/\jfour{} densitometry proves to be best suited to objects 
with \textit{T}\tsb{K} $\gtrsim$ 100 K, above which the \f{} LVG models 
become relatively independent of kinetic temperature. Compared with the similarly 
utilized \jone{} and \jtwo{} transitions, the $J=3$ and 4 \textit{K}-doublets 
provide higher spatial resolution and sensitivity to hot, dense material, which makes them 
more efficient probes of spatial density in molecular cores. However, beam widths 
comparable to the anticipated source sizes make source structure considerations 
important. Since mapping measurements have yet to be conducted for these transitions, the 
correlation between the spatial extent traced by the $J=3$ and 4 \textit{K}-doublets and 
that of other dense gas tracers is uncertain, making spatial assumptions based on past 
measurements problematic.

This work serves as a 
successful proof-of-concept for the \f{} $J=3/J=4$ \textit{K}-doublet
densitometry technique, adding a useful new diagnostic to the study of dense 
molecular environments.
	
\acknowledgements

The Green Bank Telescope (GBT) staff was characteristically 
helpful and contributed significantly to the success of our observing 
program. The authors also thank the referee for several comments that greatly 
enhanced this work. P. I. M. would like to thank Harold Butner and the NRAO 
REU students of 2009 for their continued support. Funding for this  
project was provided by the NSF through the NRAO REU Program
(award No. 0755390) and by NRAO though an undergraduate internship.
\linebreak \\
\textit{Facilities}: \facility{GBT}


\scriptsize
\begin{thebibliography}{}
\bibitem[Anthony-Twarog(1982)]{AT82} Anthony-Twarog, B. J. 1982, \aj, 87, 1213
\bibitem[Araya et al.(2006)]{Ara06} Araya, E., Hofner, P., Olmi, L., Kurtz, S., \& Linz, H. 2006, \apj, 132, 1851
\bibitem[Barsony et al.(1998)]{Bar98} Barsony, M., Ward-Thompson, D., Andr\'{e}, P., \& O'Linger, J., \apj, 1998, 509, 733
\bibitem[Bastien et al.(1985)]{Bas85} Bastien, P., Batrla, W., Henkel, C., Pauls, T., Walmsley, C. M., \& Wilson, T. L. 1985, \aap, 146, 86
\bibitem[Batrla et al.(1983)]{Bat83} Batrla, W., Wilson, T. L., Ruf, K., \& Bastien, P. 1983, \aap, 128, 279
\bibitem[Baudry \& Menten(1995)]{BM95} Baudry, A., \& Menten, K. M. 1995, \aap, 298, 905
\bibitem[Blake et al.(1987)]{Blake87} Blake, G. A., Sutton, E. C., Masson, C. R., \& Phillips, T. G. 1987, \apj, 315, 621
\bibitem[Blake et al.(1995)]{Bla95} Blake, G. A., Sandell, G., van Dishoeck, E. F., Groesbeck, T. D., Mundy, L. G, \& Aspin, C. 1995, \apj, 441, 689
\bibitem[Brogan et al.(2007)]{Bro07} Brogan, C. L., Chandler, C. J., Hunter, T. R., Shirley, Y. L., \& Sarma, A. P. 2007, \apj, 660, L133
\bibitem[Butner et al.(1991)]{But91} Butner, H. M., Evans, N. J., II, Lester, D. F., Levreault, R. M., \& Strom, S. E. 1991, \apj, 376, 636
\bibitem[Campbell et al.(1989)]{Camp89} Campbell, M. F., Lester, D. F., Harvey, P. M., \& Joy M. 1989, \apj, 345, 298
\bibitem[Chen et al.(1993)]{Chen93} Chen, H., Tokunaga, A. T., Strom, K. M., \& Hodapp, K. -W. 1993, \apj, 407, 639
\bibitem[Chen et al.(1996)]{Chen96} Chen, H., Ohashi, N., \& Umemoto, T. 1996, \aap, 112, 717
\bibitem[Churchwell et al.(1992)]{Chu92} Churchwell, E., Walmsley, C. M., \& Wood, D. O. S. 1992, \aap, 253, 541
\bibitem[Codella et al.(2005)]{Cod05} Codella, C., Bachiller, R., Benedettini, M., Caselli, P., Viti, S., \& Wakelam, V. 2005, \mnras, 361, 244
\bibitem[Cohen et al.(1984)]{Cohen84} Cohen, M., Harvey, P. M., Wilking, B. A., \& Schwartz, R. D. 1984, \apj, 278, 671
\bibitem[Comito et al.(2007)]{Com07} Comito, C., Schilke, P., Endesfelder, U., Jim\'{e}nez-Serra I., \& Mart\'{i}n-Pintado, J. 2007, \aap, 469, 207
\bibitem[Crutcher et al.(1986)]{Cru86} Crutcher, R M., Henkel, C., Wilson, T. L., Johnston, K. J., \& Bieging, J. H. 1986, \apj, 307, 302
\bibitem[Cyganowski et al.(2007)]{Cyg07} Cyganowski, C. J., Brogan, C. L., \& Hunter, T. R. 2007, \aj, 134, 346
\bibitem[Dickel \& Goss(1987)]{DG87} Dickel, H. R., \& Goss, W. M. 1987, \aap, 185, 271
\bibitem[Dickel et al.(1996)]{Dickel96} Dickel, H R., Goss, W. M., \& Condon, G. R. 1996 \apj, 460, 716
\bibitem[Di Francesco et al.(2001)]{DiF01} Di Francesco, J., Myers, P. C., Wilner, D. J., Ohashi, N., \& Mardones, D. 2001, \apj, 562, 770
\bibitem[Downes et al.(1980)]{Down80} Downes, D., Wilson, T. L., Bieging, J., \& Wink, J. 1980, \aaps, 40, 379
\bibitem[Dreher \& Welch(1981)]{DW81} Dreher, J. W., \& Welch, W. J. 1981, \apj, 245, 857
\bibitem[Eiroa et al.(2008)]{Eir08} Eiroa, C., Djupvik, A. A., \& Casali, M. M. 2008, in The Southern Sky ASP Monograph Publications, ed. B. Reipurth, Handbook of Star Forming Regions, II, 5, 693
\bibitem[Emprechtinger et al.(2009)]{Emp09} Emprechtinger, M., Wiedner, M. C., Simon, R., Wiechin, G., Volgenau, N. H., Bielau, F., Graf, U. U., G$\ddot{\mathrm{u}}$sten, R., Honingh, C. E., Jacobs, K., Rabanus, D., Stutzki, J., \& Wyrowski, F. 2009, \aap, 496, 731
\bibitem[Evans et al.(1975)]{Evans75} Evans, N. J., II, Morris, G., Sato, T., \& Zuckerman, B. 1975, \apj, 196, 433
\bibitem[Fuller et al.(1995)]{Ful95} Fuller, G. A., Ladd, E. F. Padman, R., Myers, P. C., \& Adams, F. C. 1995, \apj, 454, 862
\bibitem[Garay et al.(1996)]{Gar96} Garay, G., Ram\'{i}rez, S, Rodr\'{i}guez, L. F., Curiel, S., \& Torrelles, J. M. 1996, \apj, 459, 193
\bibitem[Garrison et al.(1975)]{Gar75} Garrison, B. J., Lester, W. A., Jr., Miller, W. H., \& Green, S. 1975, \apj, 200, L175
\bibitem[Gaume et al.(1993)]{Gau93} Gaume, R. A., Johnston, K. J., \& Wilson, T. L. 1993, \apj, 417, 645
\bibitem[Genzel \& Stutzki(1989)]{GS89} Genzel, R. \& Stutzki, J. 1989, \araa, 27, 41
\bibitem[Green(1991)]{Gre91} Green, S. 1991, \apjs, 76, 979
\bibitem[Guilloteau et al.(1983)]{GSD83} Guilloteau, S., Stier, M. T., \& Downes, D. 1983, \aap, 126, 10
\bibitem[Hasegawa \& Mitchell(1995)]{HM95} Hasegawa, T., \& Mitchell, G. F. 1995, \apj, 441, 665
\bibitem[Heaton et al.(1989)]{Heat89} Heaton, B. D., Little, L. T., \& Bishop, I. S. 1989, \aap, 213, 148
\bibitem[Helmich et al.(1994)]{Helm94} Helmich, F. P., Jansen D. J., de Graauw, Th., Groesbeck, T. D., \& van Dishoeck, E. F. 1994, \aap, 283, 626
\bibitem[Henkel et al.(1980)]{Hen80} Henkel, C., Walmsley, C. M., \& Wilson, T. L. 1980, \aap, 82, 41
\bibitem[Henkel et al.(1982)]{Hen82} Henkel, C., Wilson, T. L., \& Bieging, J. 1982, \aap, 109, 344
\bibitem[Heyer et al.(1989)]{Hey89} Heyer, M. H., Snell, R. L., Morgan, J., \& Schloerb, F. P. 1989, \apj, 346, 220
\bibitem[Hirota et al.(2008)]{Hir08} Hirota, T., Bushimata, T., Choi, Y. K., et al. 2008, \pasj, 60, 37
%Honma, M., Imai, H., Iwadate, K., Jike, T., Kameya, O., Kamohara, R., Kan-ya, Y., Kawaguchi, N., Kijima, M., Kobayashi, H., Kuji, S., Kurayama, T., Manabe, S., Miyaji, T., Nagayama, T., Nakagawa, A., Oh, C. S., Omodaka, T., Oyama, T., Sakai, S., Sasao, T., Sato, K., Shibata, K. M., Tamura, Y., \& Yamashita, K. 2008, \pasj, 60, 37
\bibitem[Hodapp \& Deane(1993)]{HD93} Hodapp, K. -W., \& Deane, J. 1993, \apjs, 88, 119
\bibitem[Hoffman et al.(2003)]{Hof03} Hoffman, I. M., Goss, W. M., Palmer, P., \& Richards, A. M. S. 2003, \apj, 598, 1061
\bibitem[Hurt et al.(1996)]{Hurt96} Hurt, R. L., Barsony, M., \& Wootten, A. 1996, \apj, 456, 686
\bibitem[Johnston et al.(1983)]{John83} Johnston, K. J., Palmer, P., Wilson, T. L., \& Bieging, J. H. 1983, \apj, 271, L89
\bibitem[Kenyon et al.(1994)]{Ken94} Kenyon, S. J., Dobrzycka, D., \& Hartmann, L. 1994, \aj, 108, 1872
\bibitem[Kim et al.(2008)]{Kim08} Kim, M. K., Hirota, T., Honma, M., et al. 2008, \pasj, 60, 991
%Kobayashi, H., Bushimata, T., Choi, Y. K., Imai, H., Iwadate, K., Jike, T., Kameya, O., Kamohara, R., Kan-ya, Y., Kawaguchi, N., Kuji, S., Kurayama, T., Manabe, Matsui, M., Matsumoto, N., Miyaji, T., Nagayama, T., Nakagawa, A., Oh, C. S., Omodaka, T., Oyama, T., Sakai, S., Sasao, T., Sato, K., Sato, M., Shibata, K. M., Tamura, Y., \& Yamashita, K. 2008, \pasj, 60, 991
\bibitem[Kitamura et al.(1992)]{Kit02} Kitamura, Y., Kawabe, R., \& Ishiguro, M. 1992, \pasj, 44, 407
\bibitem[Kuchar \& Bania(1994)]{KB94} Kuchar, T. A., \& Bania, T. M. 1994, \apj, 436, 117
\bibitem[Looney(1998)]{Loon98} Looney, L. W. 1998, Ph.D. Thesis, Univ. Maryland
\bibitem[Looney et al.(2000)]{Loon00} Looney, L. W., Mundy, L. G., \& Welch, W. J. 2000, \apj, 529, 477
\bibitem[Lugo et al.(2004)]{Lug04} Lugo, J., Lizano, S., \& Garay, G. 2004, \apj, 614, 807
\bibitem[Mangum et al.(1990)]{Man90} Mangum, J. G., Wootten, A., Wadiak, E. J., \& Loren, R. B. 1990, \apj, 348, 542
\bibitem[Mangum et al.(1991)]{Man91} Mangum, J. G., Wootten, A., \& Mundy, L. G. 1991, \apj, 378, 576
\bibitem[Mangum et al.(1992)]{Man92} Mangum, J. G., Wootten, A., \& Mundy, L. G. 1992, \apj, 388, 467
\bibitem[Mangum et al.(1993)]{Man93} Mangum, J. G., Wootten, A., \& Plambeck, R. L. 1993, \apj, 409, 282
\bibitem[Mangum \& Wootten(1993)]{MW93} Mangum, J. G., \& Wootten, A. 1993 \apjs, 89, 123
\bibitem[Mangum et al.(1999)]{Man99} Mangum, J. G., Wootten, A., \& Barsony, M. 1999, \apj, 526, 845
\bibitem[Mangum et al.(2008)]{Man08} Mangum, J. G., Darling, J., Menten, K. M., \& Henkel, C. 2008, \apj, 673, 832
\bibitem[Maret et al.(2004)]{Maret04} Maret, S., Ceccarelli C., Caux, E., Tielens, A. G. G. M., J{\o}rgenson, J. K., van Dishoeck, E, 	Bacmann, A., Castets, A., Lefloch, B., Loinard, L., Parise, B., \& Sch$\ddot{\mathrm{o}}$ier, F. L. 2004, \aap, 	416, 577
\bibitem[Martin(1972)]{Mar72} Martin, A. H. M. 1972, \mnras, 157, 31
\bibitem[Martin(1973)]{Mar73} Martin, A. H. M. 1973, \mnras, 163, 141
\bibitem[Mauersberger et al.(1988)]{Mauer88} Mauersberger, R., Wilson, T. L., \& Henkel C. 1988, \aap, 201, 123
\bibitem[McMullin et al.(1993)]{Mc93} McMullin, J. P., Mundy, L. G., \& Blake, G. A. 1993, \apj, 405, 599
\bibitem[McMullin et al.(1994)]{Mc94} McMullin, J. P., Mundy, L. G., \& Blake, G. A. 1994, \apj, 437, 305
\bibitem[McMullin et al.(2000)]{Mc00} McMullin, J. P., Mundy, L. G., Blake, G. A., Wilking, B. A., Mangum, J. G., \& Latter, W. B. 2000, \apj, 536, 845
\bibitem[Mezger et al.(1990)]{Mez90} Mezger, P. G., Zylka, R., \& Wink, J. E. 1990, \apj, 228, 95
\bibitem[Mezger et al.(1992)]{Mez92} Mezger, P. G., Sievers, A. W., Haslam, C. G. T.,  Kreysa, E., Lemke, R., Mauersberger, R., \& Wilson, T. L. 1992, \aap, 256, 631
\bibitem[Mitchell et al.(1990)]{Mit90} Mitchell, G. F., Maillard, J. -P., Allen, M., Beer, R., \& Belcourt, K. 1990, \apj, 363, 554
\bibitem[Mookerjea et al.(2007)]{Mook07} Mookerjea, B., Casper, E., Mundy, L. G., \& Looney, L. W. 2007, \apj, 659, 447
\bibitem[Moriarty-Schieven et al.(1995)]{MS95} Moriarty-Schieven, G. H., Wannier, P. G., Mangum, J. G., Tamura M., \& Olmsted, V. K. 1995, \apj, 455, 190
\bibitem[Moscadelli et al.(2010)]{Mosc10} Moscadelli, L., Xu, Y., \& Chen, X. 2010, \apj, 716, 1356
\bibitem[Myers \& Buxton(1980)]{MB80} Myers, P. C., \& Buxton, R. B. 1980, \apj, 239, 515
\bibitem[Nerf(1975)]{Nerf75} Nerf, R. B. Jr. 1975, J. Mol. Spec., 58, 451
\bibitem[Odenwald \& Schwartz(1993)]{OS93} Odenwald, S. F., \& Schwartz, P. R. 1993, \apj, 405, 706
\bibitem[Patel et al.(2005)]{Pat05} Patel, N. A., Curiel, S., Sridharan, T. K., et al. 2005, \nat, 437, 109
\bibitem[Pendleton et al.(1986)]{Pen86} Pendleton, Y., Werner, M. W., Capps, R., \& Lester, D. 1986, \apj, 311, 360
\bibitem[Persson et al.(1981)]{Per81} Persson, S. E., Geballe, T. R., Simon, T., Lonsdale, C. J., \& Baas, F. 1981, \apj, 251, L85
\bibitem[Pratap et al.(1997)]{Pra97} Pratap, P., Batrla, W., \& Snyder, L. E. 1989, \apj, 341, 832
\bibitem[Qiu et al.(2011)]{Qiu11} Qiu, K., Zhang, Q., \& Menten, K. M. 2011, \apj, 728, 6
\bibitem[Reid et al.(2009)]{Reid09} Reid, M. J., Menten, K. M., Zheng, X. W., Brunthaler, A., Moscadelli, L., Xu, Y., Zhang, B., Sato, M., Honma, M., Hirota, T., Hachisuka, K., Choi, Y. K., Moellenbrock, G. A., \& Bartkiewicz 2009, \apj, 700, 137
\bibitem[Remijan et al.(2004)]{Rem04} Remijan, A, Sutton, E. C., Snyder, L. E., Friedel, D. N., Liu, S. -Y., \& Pei, C. -C. 2004, \apj, 606, 917
\bibitem[Roberts et al.(2002)]{Rob02} Roberts, H., Fuller, G. A., Millar, T. J., Hatchell, J., \& Buckle, J. V. 2002, \aap, 381, 1026
\bibitem[Roberts \& Millar(2007)]{RM07} Roberts, H., \& Millar, T. J. 2007, \aap, 471, 849
\bibitem[Rots et al.(1981)]{Rot81} Rots, A. H., Dickel, H. R., Forster, J. R., \& Goss, W. M. 1981 245, L15
\bibitem[Rygl et al.(2010)]{Rygl10} Rygl, K. L. J., Brunthaler, A., Reid, M. J., Menten, K. M., van Langevelde, H. J., \& Xu, Y. 2010, \aap, 511, id.A2 
\bibitem[Sandell et al.(1991)]{Sand91} Sandell, G., Aspin, C., Duncan, W. D., Russell, A. P. G., \& Robson, E. I. 1991, \apj, 376, L17
\bibitem[Sandell et al.(2005)]{San05} Sandell, G., Goss, W. M., \& Wright, M. 2005, \apj, 621, 839
\bibitem[Sato et al.(2010)]{Sato10} Sato, M., Reid, M. J., Brunthaler, A., \& Menton, K. M. 2010, \apj, 720, 1055
\bibitem[Sobolev(1960)]{Sob60} Sobolev, V. V. 1960, Moving Envelopes of Stars (Cambridge: Harvard Univ. Press)
\bibitem[Stanke \& Williams(2007)]{SW07} Stanke, T., \& Williams, J. P. 2007, \apj, 133, 1307
\bibitem[Strom et al.(1989)]{Strom89} Strom, K. M., Margulis, M., \& Strom, S. E. 1989, \apj, 346, L33
\bibitem[Takakuwa et al.(2004)]{Tak04} Takakuwa, S., Ohashi, N., Ho, P. T. P., Qi, C., Wilner, D. J., Zhang, Q., Bourke, T. L., Hirano, N., Choi, M., \& Yang, J. 2004, \apj, 616, L15
\bibitem[Takano et al.(1984)]{Tak84} Takano, T., Fukui, Y., Ogawa, H., Takaba, H., Kawabe, R., Fujimoto, Y., Sugitani, K., \& Fujimoto, M. 1984, \apj, 282, L69
\bibitem[Tauber et al.(1988)]{Tau88} Tauber, J. A., Goldsmith, P. F., \& Snell, R. L. 1988, \apj, 325, 846
\bibitem[Terebey \& Padgett(1997)]{TP97} Terebey, S., \& Padgett, D. L. 1997, IAUS, 182, 507T
\bibitem[Torrelles et al.(1996)]{Tor96} Torrelles, J. M., G\'{o}mez, J. F., Rodr\'{i}guez, L. F., Curiel, S., Ho, P. T. P. \& Garay, G. 1996, \apj, 457, L107
\bibitem[Troscompt et al.(2009)]{Tro09} Troscompt, N., Faure, A., Wiesenfeld, L., Ceccarelli, C., \& Valiron, P. 2009, \aap, 493, 687
\bibitem[Turner et al.(1989)]{Tur89} Turner, B. E., Richard, L. J., \& Xu, L. 1989, ApJ, 344, 292
\bibitem[van Buren et al.(1990)]{vB90} van Buren, D., Mac Low, M. -M., Wood, D. O. S., \& Churchwell, E. 1990, \apj, 353, 570
\bibitem[van der Tak et al.(2000)]{Tak00} van der Tak, F. F. S., van Dishoeck, E. F., Evans, N. J., II., \& Blake, G. A. 2000, \apj, 537, 283
\bibitem[Volgenau et al.(2006)]{Volg06} Volgenau, N. H., Mundy, L. G., Looney, L. W., \& Welch, W. J. 2006, \apj, 651, 301
\bibitem[Walker et al.(1990)]{Wal90} Walker, C. K., Adams, F. C., \& Lada, C. J. 1990, \apj, 349, 515
\bibitem[Watanabe \& Mitchell(2008)]{WM08} Watanabe, T., \& Mitchell, G. F. 2008, \aj, 136, 1947
\bibitem[Watt \& Mundy(1999)]{WM99} Watt, S., \& Mundy, L. G. 1999, \apjs, 125, 143
\bibitem[Welch(1970)]{Wel70} Welch, W. J. 1970, \baas, 2Q, 355W
\bibitem[Werner et al.(1979)]{Wer79} Werner, M. W., Becklin, E. E., Gatley, I., Matthews, K., Neugebauer, G., \& Wynn-Williams, C. G. 1979, \mnras, 188, 463
\bibitem[Wink et al.(1994)]{Wink94} Wink, J. E., Duvert, G., Guilloteau, S., G\"{u}sten, R., Walmsley, C. M., \& Wilson, T. L. 1994, \aap, 281, 505
\bibitem[Wilson et al.(1980)]{Wil80} Wilson, T. L., Walmsley, C. M., Henkel, C., Pauls, T., \& Mattes H. 1980, \aap, 91, 36
\bibitem[Wilson et al.(1983)]{Wil83} Wilson, T. L., Walmsley, C. M., Batrla, W., \& Mauersberger, R. 1983, \aap, 127, L19
\bibitem[Wolf-Chase et al.(1998)]{Wolf98} Wolf-Chase, G. A., Barsony, M., Wootten, A., Ward-Thompson, D., Lowrance, P. J., Kastner, J. H., \& McMullin, J. P. 1998, \apj, 501, L193 
\bibitem[Wood \& Churchwell(1989)]{WC89} Wood, D. O. S., \& Churchwell, E. 1989, \apjs, 69, 831
\bibitem[Wynn-Williams et al.(1972)]{WW72} Wynn-Williams, C. G., Becklin, E. E., \& Neugebauer, G. 1972, \mnras, 160, 1
\bibitem[Zhang \& Ho(1997)]{ZH97} Zhang, Q., \& Ho, P. T. P. 1997, \apj, 488, 241
\bibitem[Zhou et al.(1990)]{Zh91} Zhou, S., Evans, N. J., II, G\"usten, R., Mundy, L. G., \& Kutner, M. L. 1991, \apj, 372, 518
\bibitem[Zylka et al.(1992)]{Zyl92} Zylka, R., G\"usten, R., Henkel, C., \& Batrla, W. 1992, \aap, 96, 525
\end{thebibliography}
\end{document}